\documentclass[11pt]{article}
\usepackage{amsfonts}
\usepackage{amsmath}
\usepackage{amsthm}
\usepackage{amssymb}
\usepackage{graphicx}
\usepackage{color}
\usepackage{placeins}
\usepackage{mathrsfs}
\usepackage{multirow}
\usepackage[top=1in, bottom=1in, left=1in, right=1in]{geometry}
\usepackage{tabularx}
\usepackage[blocks]{authblk}
\usepackage{hyperref}
\usepackage{url}
\usepackage[title]{appendix}
\usepackage{sectsty}
\usepackage{natbib}
\usepackage{setspace,caption}

\newtheorem{mydef}{Definition}
\newtheorem{myprop}{Proposition}
\setcitestyle{notesep={},round,aysep={},yysep={}}
\sectionfont{\centering\MakeUppercase}

\newcommand{\iid}{\overset{i.i.d.}{\sim}}
\allowdisplaybreaks
\input{tcilatex}
\begin{document}

\title{A Nonparametric Bayesian Approach to Copula Estimation}
\author{Shaoyang Ning \thanks{Shaoyang Ning is a PhD student at Department of Statistics, Harvard University, Cambridge, MA 02138; email: shaoyangning@fas.harvard.edu.}}
\affil{\textit{Department of Statistics, Harvard University, Cambridge, 02138, MA}
\\\url{shaoyangning@fas.harvard.edu}}
\author{Neil Shephard\thanks{Neil Shephard is a Professor of Economics and of Statistics at Department of Statistics and Department of Economics, Harvard University, Cambridge, MA 02138; email: shephard@fas.harvard.edu.}}
\affil{\textit{Department of Economics and Department of Statistics, Harvard University, Cambridge, 02138, MA}
\\\url{shephard@fas.harvard.edu}}
\date{}
\maketitle

\begin{abstract}
We propose a novel Dirichlet-based P\'olya tree (D-P tree) prior on the
copula and based on the D-P tree prior, a nonparametric Bayesian inference procedure. Through theoretical analysis and simulations, we are able to show that
the flexibility of the D-P tree prior ensures its consistency in copula
estimation, thus able to detect more subtle and complex copula structures
than earlier nonparametric Bayesian models, such as a Gaussian copula
mixture. Further, the continuity of the imposed
D-P tree prior leads to a more favorable smoothing effect in copula
estimation over classic frequentist methods, especially with small sets of
observations. We also apply our method to the copula prediction
between the S\&P 500 index and the IBM stock prices during the 2007-08
financial crisis, finding that D-P tree-based methods enjoy strong
robustness and flexibility over classic methods under such irregular market
behaviors.

KEY WORDS: copula, P\'olya tree, nonparametric Bayes, Gaussian copula
mixture model, kernel method
\end{abstract}

\affil{\textit{Department of Statistics, Harvard University, Cambridge, 02138, MA}
\\\url{shaoyangning@fas.harvard.edu}}

\affil{\textit{Department of Economics and Department of Statistics, Harvard University, Cambridge, 02138, MA}
\\\url{shephard@fas.harvard.edu}}

\baselineskip=20pt

\section{Introduction}

The copula, as the ``link" of a multivariate distribution to its marginals,
has attracted growing interest in statistical research since \cite%
{sklar1959fonctions}. By Sklar's Theorem, a copula characterizes the
dependence structure between the marginal components. Therefore, the copula
plays a central role in multivariate studies and has gained increasing
popularity in application to fields such as risk analysis, insurance
modeling, and hydrologic engineering 
\citep{nelsen2007introduction,
wu2013bayesian}.

The estimation of copulas has been well studied in parametric and
semi-parametric settings, but little work has been released on the
nonparametric Bayesian inference. In this article, we propose a novel
multi-partition Dirichlet-based P\'olya tree (D-P tree) prior on the copula.
Our D-P tree prior relaxes the binary partition constraints on earlier
P\'olya-tree-like priors but still preserves the favorable properties of the
P\'olya tree, including conjugacy and absolute continuity. Based on such a
D-P tree prior, we provide a nonparametric Bayesian approach for copula
estimation. Its consistency is validated through theoretical analysis. 

The D-P tree prior overcomes the severe bias problem of previously proposed
P\'olya-tree-like priors, and the inconsistency issue of family-based
nonparametric Bayesian approaches such as the Gaussian copula mixture %
\citep{dortetbayesian} under model misspecification. Further, compared with
classic nonparametric frequentist methods, including the empirical copula
estimation and the kernel method, the D-P tree shows a more favorable
smoothing effect, especially based on small sets of observations. We
illustrate our new method by focusing on copula structure prediction between
the S\&P 500 daily index and the IBM daily stock prices during the 2007-08
financial crisis. We find that D-P tree-based methods are rather robust and
adaptive to irregular market behavior, especially in comparison with
commonly-adopted parametric models and the empirical method.

Earlier parametric or semi-parametric methods often model copula functions
within certain parametric copula families and estimate the parameters by
maximum likelihood (ML). For marginals, either parametric or nonparametric
estimations are usually adopted 
\citep{joe1997multivariate,
jaworski2010copula, chen2007nonparametric, oakes1982model,
oakes1986semiparametric, genest1995semiparametric}. However, these
parametric or semi-parametric methods suffer from the risk of severe bias
when the model is misspecified, thus lack the flexibility to provide
accurate estimation for more complex and subtle copula structures. In
addition, copula itself is strictly-increasing-transform invariant %
\citep{schweizer1981nonparametric}. Thereby, under no further parametric
assumptions, the rank statistics of data would preserve sufficient
information required for the estimation. In light of these features,
nonparametric methods seem to be more natural and coherent for the
estimation of copula.

Most of the recent studies on nonparametric copula estimation focus on
empirical methods \citep{jaworski2010copula, deheuvels1979fonction}, or
kernel-related methods 
\citep{charpentier2007estimation, behnen1985rank,
gijbels1990estimating, schuster1985incorporating, hominal1979estimation,
devroye1985nonparametric, gasser1979kernel, john1984boundary,
muller1991smooth, chen2007nonparametric}. Current nonparametric Bayesian
methods focus mainly on an infinite mixture of elliptical copula families
such as the Gaussian or the skew-normal \citep{wu2013bayesian}. Yet such
models still have limitations: a heavy computational burden as they are
implemented through MCMC, and an inconsistency when the model is
misspecified, taking the infinite Gaussian copula mixture for a non-symmetric target copula as an instance. These motivate us to explore priors with conjugacy and more generality.

Note that here we focus mainly on the bivariate copula case to illustrate
our method, and we will discuss higher-dimensional cases towards the end.
Also, to concentrate on the estimation of copula structures itself, we assume
that the marginals are known or can be accurately estimated. So
equivalently,  in our simulations, we are concerned mainly with marginally uniform data generated
from copula distributions. Such an assumption is
reasonable in that: (1) usually we have more information (either parametric
or nonparametric) on the marginals of the data for the estimation; (2)
multivariate data are exponentially enriched when considered marginally,
providing higher resolution for accurate estimation. Yet we will discuss the
scenarios where marginal distributions are to be empirically estimated.

The article is organized as follows: in Section \ref{sec:PT}, we establish
some notation and review previous attempts for copula estimation based on
the P\'olya tree prior and their limitations. In Section \ref{sec:DPT}, we
introduce the proposed D-P tree prior and the procedure for copula
inference. In Section \ref{sec: DPTprop}, we elaborate on properties of the
D-P tree. Section \ref{sec: simu} provides a simulation-based evaluation of
our method in comparison with other common copula estimation methods. In
Section \ref{sec: app}, we provide an application of our method to the
analysis of a bivariate stock-index copula structure. We discuss the copula
estimation with unknown marginal distributions and the higher-dimensional
cases in Section \ref{sec: diss}. Section \ref{sec: concld} concludes the
article.

\section{The quasi-P\'olya tree prior on copula}

\label{sec:PT}

\subsection{The P\'olya Tree Prior}

Our focus here is on finding P\'olya-tree-like priors placed on a copula.
The P\'olya tree (PT) prior is a tractable case of a tail-free process %
\citep{ferguson1974prior}, which also includes the Dirichlet process (DP) as
a special case. But unlike the Dirichlet process, the P\'olya tree delivers
absolutely continuous measures with probability one by certain choices of
the hyper-parameters, which is the attraction for our applications.

Following the definition by \cite{lavine1992some} (Appendix \ref{append:def
copula}), suppose we have a probability measure $\mathcal{P}$ that follows a
P\'olya tree prior, i.e., $\mathcal{P}\sim PT(\Pi,\mathcal{A})$. The conjugacy of the P\'olya tree follows in that, with one observation $Y|%
\mathcal{P}\sim\mathcal{P}$, the posterior $\mathcal{P}|Y$ still follows a
P\'olya tree distribution denoted by $PT(\Pi,\mathcal{A}|Y)$ with the
hyper-parameters updated by 
\begin{align}
\alpha_\epsilon|Y=%
\begin{cases}
\alpha_\epsilon+1 & \text{if } \,Y\in B_\epsilon, \\ 
\alpha_\epsilon & \text{otherwise}.%
\end{cases}%
\end{align}

In practice, to ensure the absolute continuity of measures given by the
P\'olya tree prior, the hyper-parameters usually take as $%
\alpha_\epsilon=zm^2$ at $m$-th level of the partition, where $z$ is a fixed
constant, and the infinite-level P\'olya tree is approximated by
terminating the sampling process from $PT(\Pi,\mathcal{A})$ at finite level $%
M$.

Therefore, the PT can be intuitively viewed as a smoothed random histogram,
and enjoys favorable features such as conjugacy and absolute continuity.
Note that \cite{hanson2012inference} studied the finite mixture of P\'olya trees; \cite{paddock2003randomized} and \cite{wong2010optional} extended the classic PT with randomized partitions to
embraces higher flexibility; \cite{filippi2016bayesian} applied P\'olya tree to independence test based on the Bayes factor. So it seems promising to start with the PT
in search of a more favorable nonparametric prior for Bayesian copula
inference.

\subsection{Dortet-Bernadet's quasi-P\'olya tree prior on copula}

\label{sec: quasi polya} 
\begin{figure}[!h]
\centering
\FloatBarrier
\includegraphics[width=6.5in]{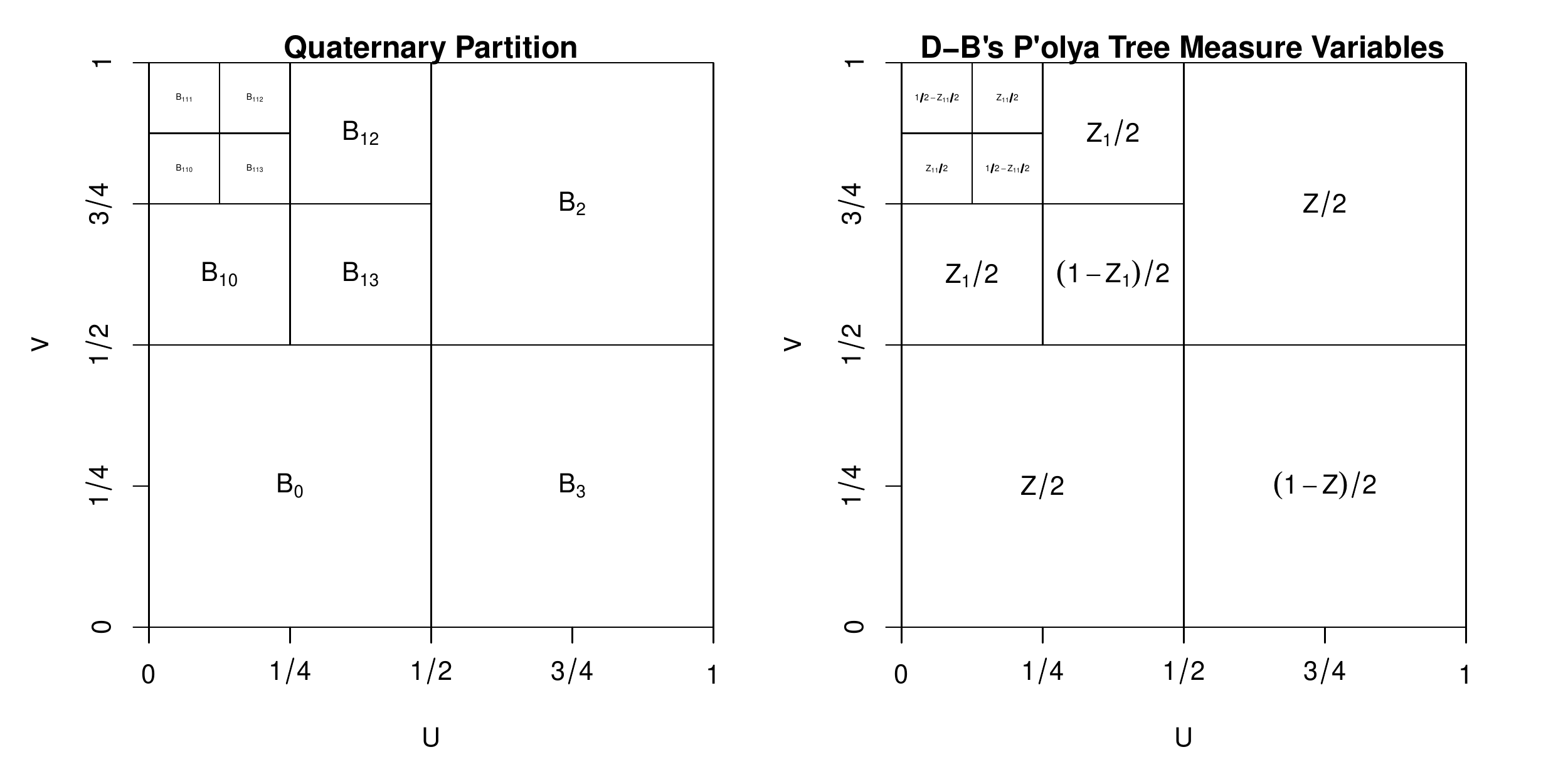}
\caption{The quaternary partition (left) on the support $[0,1]^2$ of a
bivariate copula and the parametrization of Dortet-Bernadet's quasi-P\'olya
tree prior (right).}
\label{fig:dbpt}
\end{figure}

To our knowledge, \cite{dortetbayesian} made the
first attempt to apply the PT prior to the inference of a bivariate copula
on $I=[0,1]\times[0,1]$. At each level, each square partition $B_\epsilon$ is
split into four sub-partitions $\{B_{\epsilon0},B_{\epsilon1},B_{%
\epsilon2},B_{\epsilon3}\}$ by dyadic partitions on its margins. Thereby, a
partition of $I$ is obtained by $\Pi=\{B_\epsilon\}$, $\epsilon\in\{%
\emptyset,0,1,2,3,00,01,02,03\dots\}$, demonstrated by the left panel of
Figure \ref{fig:dbpt}. 

The P\'{o}lya-tree-like probability measure $\mathcal{P}$, which we call a
quasi-P\'{o}lya tree prior, is defined by independent variables $\mathcal{Z}%
=\{Z_{\epsilon }\}$, hyper-parameters $\mathcal{A}=\{\alpha _{\epsilon
0},\alpha _{\epsilon 1}\}$, where $Z_{\epsilon }\sim Beta(\alpha _{\epsilon
0},\alpha _{\epsilon 1})$ and 
\begin{equation*}
\mathcal{P}(B_{\epsilon =\epsilon _{1}\epsilon _{2}\dots \epsilon
_{m}})=\left( \prod_{j=1;\epsilon _{j}=0\text{ or }\epsilon
_{j}=2}^{m}Z_{\epsilon _{1}\epsilon _{2}\dots \epsilon _{j-1}}/2\right)
\left\{ \prod_{j=1;\epsilon _{j}=1\text{ or }\epsilon
_{j}=3}^{m}(1-Z_{\epsilon _{1}\epsilon _{2}\dots \epsilon _{j-1}})/2\right\}
.
\end{equation*}%
The posterior-like hyper parameters are updated as: 
\begin{align}
\alpha _{\epsilon 0}|Y& =%
\begin{cases}
\alpha _{\epsilon 0}+1 & \text{if }\,Y\in B_{\epsilon 0}\cup B_{\epsilon 2},
\\ 
\alpha _{\epsilon 0} & \text{otherwise};%
\end{cases}
&\alpha _{\epsilon 1}|Y =%
\begin{cases}
\alpha _{\epsilon 1}+1 & \text{if }\,Y\in B_{\epsilon 1}\cup B_{\epsilon 3},
\\ 
\alpha _{\epsilon 1} & \text{otherwise}.%
\end{cases}%
\end{align}

Unfortunately, Dortet-Bernadet's quasi-P\'olya tree prior performs rather
unsatisfactorily even in simple bivariate Gaussian copula case. As shown in Figure \ref%
{fig: dbpt dpt}, where we estimate the Gaussian copula with $\rho=0.9$ based
on $N=10,000$ data points and approximation level $M=10$, the ``grid" effect
is severe for such a quasi-P\'olya Tree prior, leading to considerable bias
for estimation. 
\begin{figure}[!h]
\centering
\FloatBarrier
\includegraphics[width=6.5in]{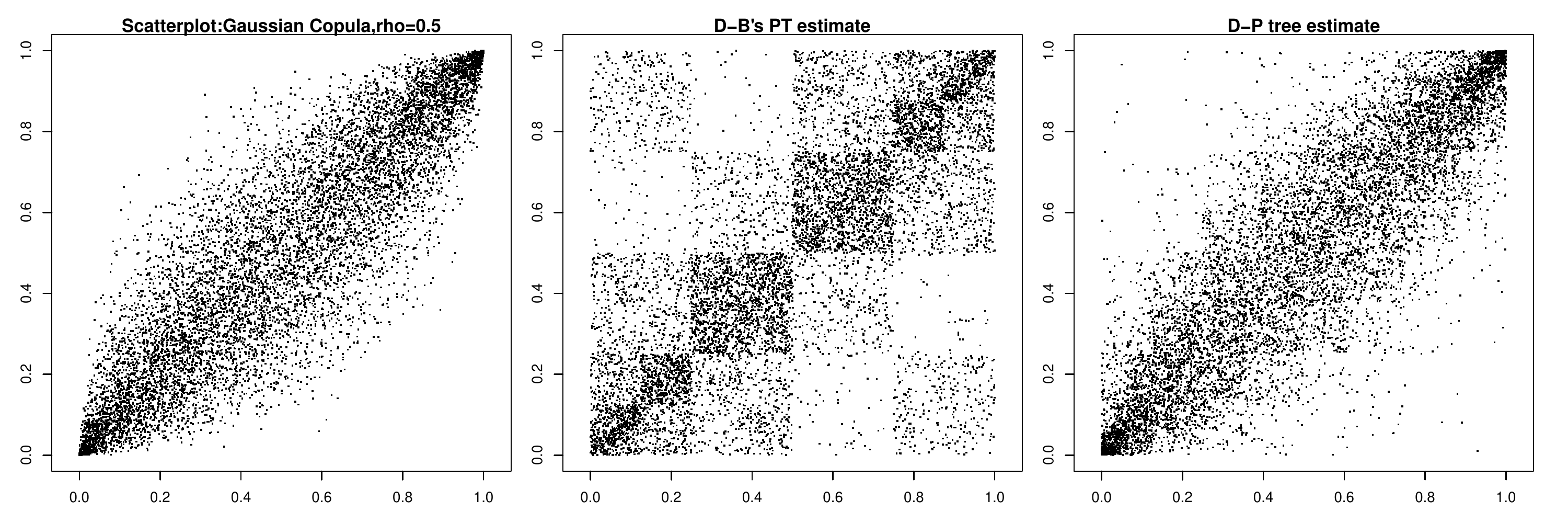}
\caption{Scatterplots comparing the Gaussian copula estimates: the Gaussian
copula (left), the quasi-P\'olya tree (middle) and the D-P tree (right)
priors.}
\label{fig: dbpt dpt}
\end{figure}

In fact, Dortet-Bernadet's P\'olya tree prior deviates from the classic
P\'olya tree in that it mixes a binary partition with a quaternary partition
across levels. It does not preserve the features of PT such as the
conjugacy, so the posterior-like update is rather ad hoc. Further, it puts
strong constraints on its dependence structure by combining the two diagonal
dyadic sub-partitions at each level when updating the hyper-parameters for
posterior, which causes severe bias when the true copula is heavily
asymmetric in the super-partition at the previous level.

\section{Our approach: Dirichelet-based P\'{o}lya tree}

\label{sec:DPT}

\subsection{The Dirichlet-based P\'olya Tree (D-P tree)}
\label{subsec:dptree}
One natural way to remedy the inflexibility in the design of the quasi-P\'olya tree is to adopt the more flexible Dirichlet distribution
for measure variables ($Z_\epsilon$) in place of the much-constrained Beta
distribution in the classic PT. Here we first give the Dirichlet-based
P\'olya tree a general definition:
%%definition for tree measurable partitions

\begin{mydef}
Let $\Omega$ be a separable measurable space. We say a partition $%
\Pi=\{B_\epsilon\}$ of $\Omega$ is one of its measurable tree partitions if

\begin{itemize}
\item the subpartitions at level $m+1$ $\{B_{\epsilon _{1}\dots \epsilon
_{m+1}}\}$ is refinement of previous level $\{B_{\epsilon _{1}\dots \epsilon
_{m}}\}$;

\item $\Pi=\{B_\epsilon\}$ generates measurable sets of $\Omega$.
\end{itemize}
\end{mydef}

%%definition for D-P tree

\begin{mydef}
\label{def:dpt}Let $\Omega $ be a separable measurable space and $\Pi
=\{B_{\epsilon }\}$ be one of its measurable tree partitions. A random
probability measure $\mathcal{P}$ is said to have a Dirichlet-based P\'{o}%
lya tree distribution, or D-P tree prior, with parameters ($\Pi $,$\mathcal{A%
}$), written $\mathcal{P}\sim DPT(\Pi ,\mathcal{A})$, if there exists
non-negative numbers $\mathcal{A}=\{\alpha _{\epsilon }\}$ and random
variables $\mathcal{Z}=\{\boldsymbol{Z}_{\epsilon }\}$ such that the
following hold:

\begin{itemize}
\item all the random vectors in $\mathcal{Z}$ are independent;

\item for every $m=1,2,\dots $ and every sequence $\epsilon =\epsilon
_{1}\epsilon _{2}\dots \epsilon _{m}$, $\boldsymbol{Z}_{\epsilon
}=(Z_{\epsilon 0},\dots ,Z_{\epsilon k_{\epsilon }})\sim Dirichlet(\alpha
_{\epsilon 0},\dots ,\alpha _{\epsilon k_{\epsilon }})$, with $B_{\epsilon
}=\cup _{i=0}^{k_{\epsilon }}B_{\epsilon i}$ and $k_{\epsilon }$ the number
of subpartitions in $B_{\epsilon }$;

\item for every $\epsilon $, 
$\mathcal{P}(B_{\epsilon =\epsilon _{1}\epsilon _{2}\dots \epsilon
_{m}})=\left( \prod_{j=1}^{m}Z_{\epsilon _{1}\epsilon _{2}\dots \epsilon
_{j}}\right)$.
\end{itemize}
\end{mydef}

The D-P tree prior still falls into the general class of tail-free process,
as the random variables for measures are independent across different
partition levels. Yet rather than constraining on binary partitions and beta
distributions, the D-P tree adopts a more flexible partition structure and,
accordingly, the Dirichlet-distributed variables for the measures, which
preserves similar properties to the classic P\'olya tree prior.

\subsection{Conjugacy and Posterior Updating}

\begin{figure}[!h]
\centering
\FloatBarrier
\includegraphics[width=6.5in]{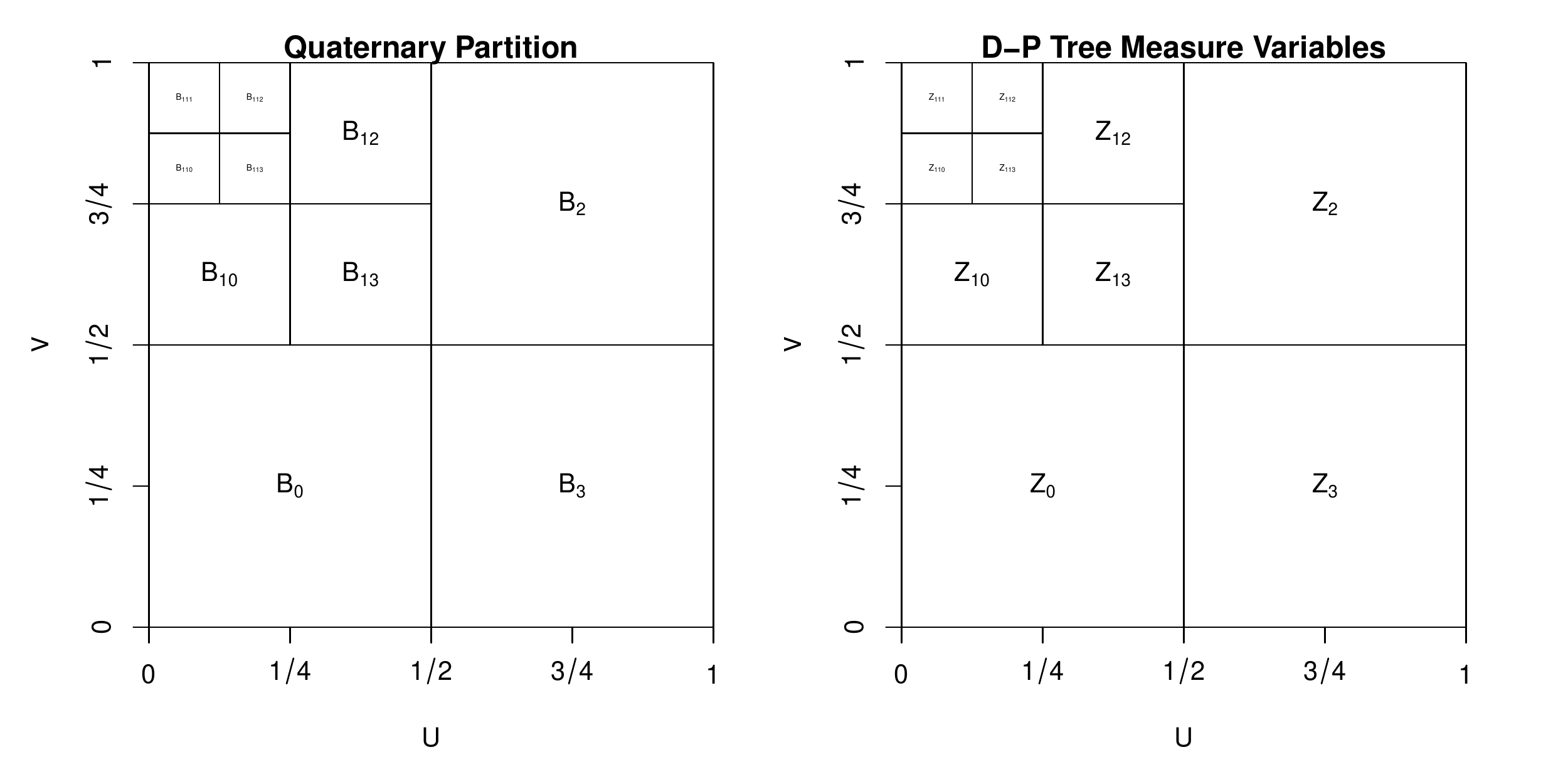}
\caption{The quaternary partition (left) on the support $[0,1]^2$ of a
bivariate copula and the parametrization of Dirichelet-based tree (D-P tree)
prior (right).}
\label{fig:dpt}
\end{figure}

Adapting the D-P tree prior to bivariate copula estimation, we constrain the
D-P tree on $\Omega=I=[0,1]\times[0,1]$, with the quaternary dyadic
partition $\Pi=\{B_{\epsilon0},B_{\epsilon1},B_{\epsilon2},B_{\epsilon3}\}$,
which repeats Section \ref{sec: quasi polya}, but now the hyper-parameters $%
\mathcal{A}=\{\alpha_{\epsilon0},\alpha_{\epsilon1},\alpha_{\epsilon2},%
\alpha_{\epsilon3}\}$ and random variables $(Z_{\epsilon0},Z_{\epsilon1},Z_{%
\epsilon2},Z_{\epsilon3})\sim
Dirichlet(\alpha_{\epsilon0},\alpha_{\epsilon1},\alpha_{\epsilon2},\alpha_{%
\epsilon3})$, as illustrated in Figure \ref{fig:dpt}. From now on, without
further specification, we focus only on the D-P tree prior with such a
quaternary dyadic partition parametrization, though all results can be
generalized.

Such D-P tree prior preserves the conjugacy property of original P\'olya
tree, thus with $\mathcal{P}\sim DPT(\Pi,\mathcal{A})$ and an observation $Y|%
\mathcal{P}\sim \mathcal{P}$, the posterior $\mathcal{P}|Y$ can be readily
updated.

\begin{myprop}[Conjugacy]
\label{prop: dpt conjugacy} Let $\mathcal{P}$ be a measure on $I=[0,1]\times
\lbrack 0,1]$, and an observation $Y|\mathcal{P}\sim \mathcal{P}$. Suppose $%
\mathcal{P}$ follows a D-P tree prior, as $\mathcal{P}\sim DPT(\Pi ,%
\mathcal{A})$, with the quaternary partition $\Pi =\{B_{\epsilon }\}$ and
Dirichlet-distributed random variables $\mathcal{Z}=\{Z_{\epsilon }\}$ and
hyper-parameters $\mathcal{A}=\{\alpha _{\epsilon 0},\alpha _{\epsilon
1},\alpha _{\epsilon 2},\alpha _{\epsilon 3}\}$. Then the posterior $%
\mathcal{P}|Y\sim DPT(\Pi ,\mathcal{A}|Y)$, where, for $i=0,1,2,3$, 
\begin{equation*}
\alpha _{\epsilon i}|Y=%
\begin{cases}
\alpha _{\epsilon i}+1 & \text{{\normalfont if} }\,Y\in B_{\epsilon i}, \\ 
\alpha _{\epsilon i} & \text{{\normalfont otherwise}}.%
\end{cases}%
\end{equation*}
\end{myprop}

Proof: 
$p(\mathcal{Z}|Y)\propto p(Y|\mathcal{Z})p(\mathcal{Z})\propto
\prod_{j=1}^{\infty }Z_{\epsilon _{1}\dots \epsilon _{j}}\prod Z_{\epsilon
}^{\alpha _{\epsilon }}\propto \prod Z_{\epsilon }^{\alpha _{\epsilon
}+I_{Y\in B_{\epsilon }}} \square$.

For $N$ i.i.d. observations $\boldsymbol{Y}=(Y_1, Y_2, \dots, Y_N)$, the
posterior update for multiple observations is rather intuitive and
straightforward: at each level of the partitions, the hyper-parameter $%
\alpha_{\epsilon}$ associated with the specific partition $B_{\epsilon}$ is
incremented by the number of observations falling in that partition, denoted
by $n_\epsilon$, where $n_\epsilon=\sum_{i=1}^N I_{Y_i\in B_{\epsilon}}$.
Simply put: $\alpha_\epsilon|\boldsymbol{Y}=\alpha_\epsilon+n_\epsilon$.

\subsection{Copula Estimation by the D-P Tree Prior}

\label{subset: dpt copula} For the copula estimation, suppose we have $N$
i.i.d. observations $\boldsymbol{Y}=(Y_1, Y_2, \dots, Y_N)$ from an unknown
copula distribution $C$, i.e.,$Y_1, Y_2, \dots, Y_N \overset{i.i.d.}{\sim} C$. We assume that $C$ follows a D-P tree prior, i.e., $C\sim DPT(\Pi, \mathcal{A%
})$, where we take $\Pi$ to be the quaternary partition on the unit square $%
[0,1]\times[0,1]$ and $\mathcal{A}=\{\alpha_\epsilon:\alpha_{\epsilon_1\dots%
\epsilon_m}=m^2\}$. By Proposition \ref{prop: dpt conjugacy}, the posterior $%
C|\boldsymbol{Y}\sim DPT(\Pi, \mathcal{A}|\boldsymbol{Y})$, where $\mathcal{A%
}|\boldsymbol{Y}=\{\alpha:
\alpha_{\epsilon_1\dots\epsilon_m}=m^2+n_\epsilon\}$.

Therefore, the D-P tree posterior on copula strongly resembles the
construction of a histogram of the observations, but regularized by the
imposed prior. Later we will show the choice of hyper-parameters, as in $%
\mathcal{P}\sim DPT(\Pi, \mathcal{A}=\{\alpha_\epsilon:
\alpha_{\epsilon_1\dots\epsilon_m}=m^2\})$, ensures generating absolutely
continuous measures centered on the uniform distribution, and thus the
posterior then can be viewed as a shrunk version of the histogram.

In practice, we approximate the infinite-level D-P tree prior with its $M$%
-level approximation $\mathcal{P}$:

\begin{mydef}
For a probability measure $\mathcal{P}$ such that $\mathcal{P}\sim DPT(\Pi ,%
\mathcal{A})$, with the same notation as in Definition \ref{def:dpt}, its $%
M $-level approximation $\mathcal{P}_{M}$ is, for any measurable set $B\in
\{B_{\epsilon =\epsilon _{1}\epsilon _{2}\dots \epsilon _{M}}\}$, 
\begin{equation*}
\mathcal{P}_{M}(B)=\left( \prod_{j=1}^{M}Z_{\epsilon _{1}\epsilon _{2}\dots
\epsilon _{j}}\right) \frac{\mu (B)}{\mu (B_{\epsilon =\epsilon _{1}\epsilon
_{2}\dots \epsilon _{M}})},
\end{equation*}%
where $\mu $ is the uniform measure on $\Pi $.
\end{mydef}

\section{Properties of D-P tree}

\label{sec: DPTprop}

\subsection{Equivalence to the P\'olya Tree}
We first show that, through a re-parametrization, the D-P tree prior on the unit square with the quaternary partition complies with a
classic P\'olya tree by sequentially combining the quaternary partitions to
binary partitions.

\begin{myprop}[Equivalence to the P\'olya tree]
Given a D-P tree prior on $I=[0,1]\times[0,1]$ with the quaternary partition 
$\Pi=\{B_\epsilon\}$ and Dirichlet-distributed random variables $\mathcal{Z}%
=\{Z_\epsilon\}$ and hyper-parameters $\mathcal{A}=\{\alpha_{\epsilon0},%
\alpha_{\epsilon1},\alpha_{\epsilon2},\alpha_{\epsilon3}\}$, an equivalent
P\'olya tree prior with binary partition $\tilde{\Pi}=\{\tilde{B}_\eta\}$
and Beta-distributed $\mathcal{\tilde{Z}}=\{\tilde{Z}_\eta\}$ and
hyper-parameters $\mathcal{\tilde{A}}=\{\tilde{\alpha}_{\eta0},\tilde{\alpha}%
_{\eta1}\}$ can be constructed as 
\begin{align*}
\tilde{B}_{\eta_1\eta_2\dots\eta_{2k}}=B_{\epsilon_1\epsilon_2\dots%
\epsilon_{k}},\,\,
\tilde{B}_{\eta_1\eta_2\dots\eta_{2k+1}}=B_{\epsilon_1\epsilon_2\dots%
\epsilon_{k}(2\eta_{2k+1})}\cup
B_{\epsilon_1\epsilon_2\dots\epsilon_{k}(2\eta_{2k+1}+1)},\\
\tilde{\alpha}_{\eta_1\eta_2\dots\eta_{2k}}=\alpha_{\epsilon_1\epsilon_2%
\dots\epsilon_{k}},\,\,
\tilde{\alpha}_{\eta_1\eta_2\dots\eta_{2k+1}}=\alpha_{\epsilon_1\epsilon_2%
\dots\epsilon_{k}(2\eta_{2k+1})}+\alpha_{\epsilon_1\epsilon_2\dots%
\epsilon_{k}(2\eta_{2k+1}+1)}
\end{align*}
where $k=0,1,\dots$, $\epsilon_{i}=2\eta_{2i-1}+\eta_{2i}$, $i=1,2,\dots k$.
\end{myprop}

This result follows directly from the property of representing a Dirichlet
distribution by independent Gamma distributions, and the independence
property between the represented Beta and Gamma distributions. With such
equivalence, some of the favorable features of the classic P\'olya tree
prior can be naturally extended to the D-P tree.

\subsection{Continuity of D-P Tree Prior}

Here we show that the D-P tree prior inherits the feature of generating
absolute continuous probability measures under certain constraints on the
hyper-parameters $\mathcal{A}$.

\begin{myprop}[Absolute continuity]
A D-P tree prior on $I=[0,1]\times[0,1]$ with the quaternary partition $%
\Pi=\{B_\epsilon\}$ and Dirichlet-distributed random variables $\mathcal{Z}%
=\{Z_\epsilon\}$ and hyper-parameters $\mathcal{A}=\{\alpha_{\epsilon0},%
\alpha_{\epsilon1},\alpha_{\epsilon2},\alpha_{\epsilon3}\}$ generates an
absolute continuous probability measure on $I$ with probability one when
hyper-parameters on the m-level $\alpha_{\epsilon_1\dots\epsilon_m}\propto
O(m^{1+\delta})$, $\delta>0$.

Further, with $\boldsymbol{Y}=(Y_1, Y_2, \dots, Y_N)|P \overset{i.i.d.}{\sim}
\mathcal{P}$, $\mathcal{P}\sim DPT(\Pi, \mathcal{A})$, the posterior $%
DPT(\Pi, \mathcal{A}|\boldsymbol{Y})$ also generates an absolute continuous
probability measure with probability one.
\end{myprop}

The results follow from Theorem 1.121 and Lemma 1.124 in %
\citep{schervish1995theory}. Thereby, as we implied earlier in Section \ref%
{subset: dpt copula}, the canonical hyper-parameter choice, i.e., $%
\alpha_{\epsilon_1\dots\epsilon_m}=m^2$ will lead to a D-P tree prior
that yields absolutely continuous random probability measures, which justifies
the smoothing effect of the D-P tree prior in copula estimation.

%%%%%%%%%%%%
%%Asymptotics

\subsection{Consistency of the D-P Tree Posterior}

\label{asymp} Suppose we have $N$ i.i.d. observations $\boldsymbol{Y}%
=\{Y_{1},\dots ,Y_{N}\}$ generated from true copula distribution $C$. For copula estimation, we assume $Y_i|C\iid \mathcal{C}$, with a D-P tree prior ${C}\sim DPT(\Pi,\mathcal{A}\})$. Let $\mathcal{P}_{M}$ be the M-level approximation of $C$
and $\mathcal{A}$ be canonical, i.e., the m-level hyper-parameter $%
\alpha _{\epsilon _{1}\dots \epsilon _{m}}=m^{2}$. 

For the approximated posterior $\mathcal{P}_{M}|\boldsymbol{Y}$, we have the point-wise convergence to the target copula distribution
in terms of any measurable set in the unit square:

\begin{myprop}[Point-wise convergence]
\label{prop: pw convg} For any measurable set $B\subset I=[0,1]\times
\lbrack 0,1]$, with $N\propto O(M^{3+\eta })$, $\eta >0$, then  $\text{E}((%
\mathcal{P}_{M}(B)|\boldsymbol{Y})-C(B))\rightarrow 0$, $\text{var}(\mathcal{%
P}_{M}(B)|\boldsymbol{Y})=O(\frac{M}{N})$, therefore $\mathcal{P}_{M}(B)|%
\boldsymbol{Y}\overset{p}{\rightarrow }C(B)$.
\end{myprop}

Notice that here we require that the sample size goes to infinity with a
higher order than $O(M^{3})$, which leaves the variance of our D-P posterior ($O(\frac{M}{N})$)
in a higher order than the empirical copula estimator ($O(\frac{1}{N})$). In fact, by introducing such a D-P tree prior, we sacrifice some
asymptotic statistical efficiency in exchange of the continuity of our
estimator. Also, as shown in the simulation results later, we gain
some advantages in prediction precision with small sets of observations.

If we put smoothness constraints on the target distribution, we can have
similar convergence results uniformly on $I$ for the posterior, and further
the consistency of the posterior.

\begin{myprop}[Consistency]
\label{prop:unif convg} If $C\in C^{1}([0,1]\times \lbrack 0,1])$, for $%
B\subset I$ measurable, 
$\sup_{B}|\text{E}(\mathcal{P(B)}_{M}|\boldsymbol{Y})-C| =\max \{O\left( 
\frac{M}{\sqrt{N}\gamma (M)}\right) ,O\left( \frac{M^{3}}{N\gamma (M)}%
\right) \}$;  $\sup_{B}\text{var}(\mathcal{P(B)}_{M}|\boldsymbol{Y}) =O\left( \frac{M}{%
N\gamma (M)}\right) 
$, where $\gamma (M)\sim \min_{C(B_{M})>0}C(B_{M})$. 

Further, with $N\propto O({%
2^{10M}M^{2+\eta }})$, $\eta >0$, $\forall \delta >0$ as $M\rightarrow
\infty $, $P(d_{TV}(\mathcal{P}_{M},C)\geq \delta |Y)\rightarrow 0$.
Note that $d_{TV}$ is the total variation distance between probability
measures. % and 
\end{myprop}

Specifically, we refine the order of convergence for several classic copula
distributions, which, in practice, may serve as general guidance for the
choice of partition level $M$ based on sample size $N$.

\begin{myprop}
\label{prop: order} The order requirement for the uniform convergence of
specific target copulas:

\begin{enumerate}
\item For a lower-bounded copula density, i.e., $c\geq \xi >0$, $%
\gamma(M)\geq 2^{-2M}\xi$, thus $N\propto O(M^{2+\eta}2^{4M})$;

\item For a bivariate Gaussian copula, $\gamma(M)\geq \Phi^2(\sqrt{1-|\rho|}
\Phi^{-1}(2^{-M}))\sqrt{\frac{1-|\rho|}{1+|\rho|}}$, thus $%
N\propto O(M^{2+\eta}2^{4M})$.
\end{enumerate}
\end{myprop}

Proofs of these results are provided in Appendix \ref{append:prf}. Such
convergence properties ensure the consistency of the estimation based on the
D-P tree prior, giving the D-P tree prior advantages over family-based
estimation methods under model misspecification.

%%%%%%%%%
%%simulation

\section{Simulation experiments}

\label{sec: simu} %%gaussian copula

\subsection{Evaluation: Common Copulas}

To evaluate the performance of our copula estimation procedure, we
conduct simulation studies based on common copulas as listed in Supplementary Material %
\ref{append:copula} with various parameter settings, among which Gaussian,
Student's and Gumbel are symmetric while the skew-normal is asymmetric.

For each simulation, the procedure is as follows: we first draw i.i.d. data
samples from true copula $C$ with the size of $N$, denoted by $\boldsymbol{Y}
$; then we follow the procedure described in Section \ref{subset: dpt copula}
for the posterior inference on $C$; once posterior $DPT(\Pi,\mathcal{A}|%
\boldsymbol{Y})$ is obtained, we draw 10,000 posterior predictive samples
from $\mathcal{P}_M|\boldsymbol{Y}$ to plot the scatterplots, shown in Figure \ref%
{nsim1000}-\ref{nsim100000}. Note that without further clarification, all
simulations are done with approximation level $M=10$.

\subsubsection{Posterior Scatterplots}

\label{scplot} 
\begin{figure}[!h]
\centering
\includegraphics[width=6.6in]{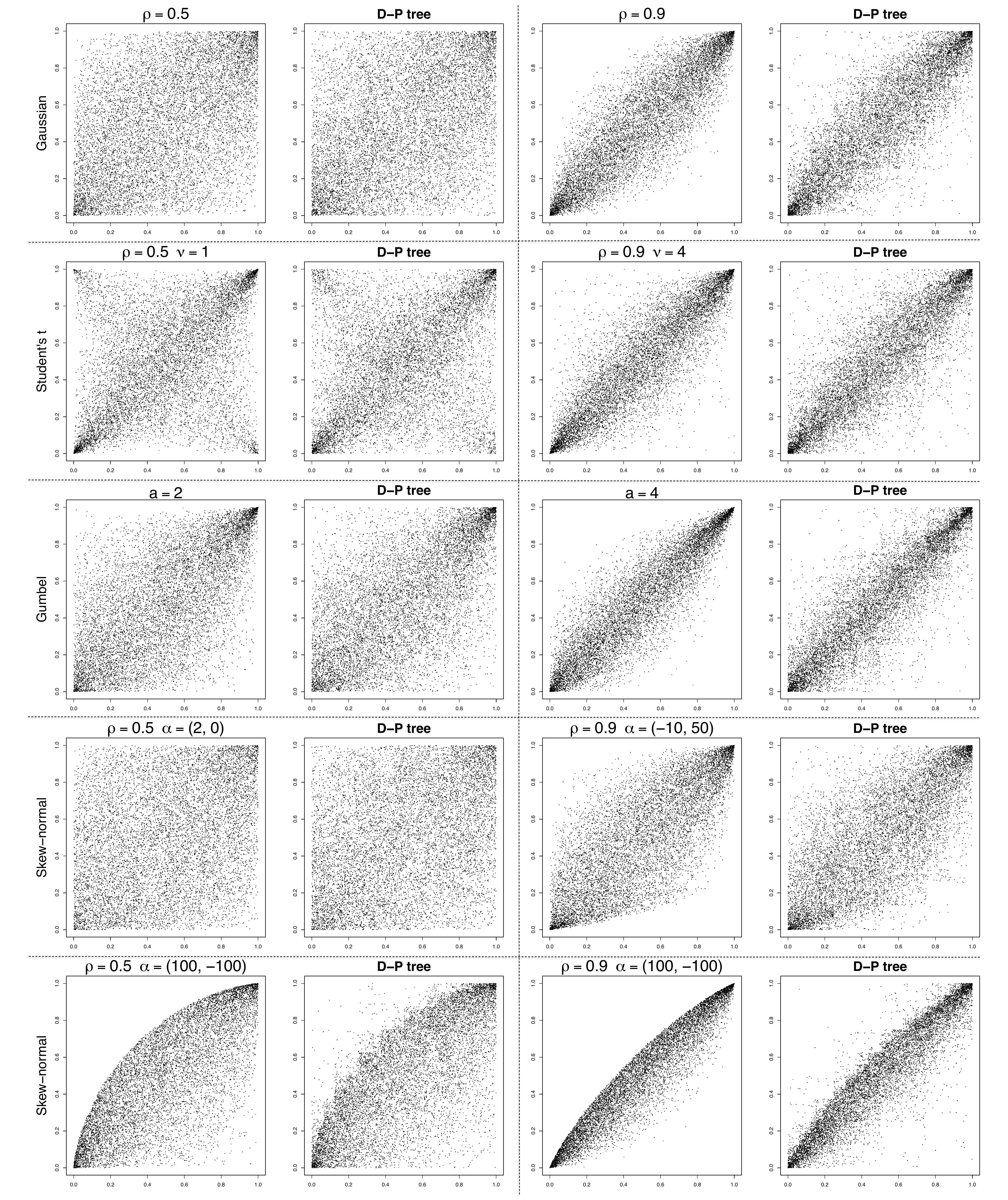}
\caption{Scatterplots of i.i.d. draws from the true copula distribution
(left) vs. the D-P tree posterior (right): sample size $N=10,000$, partition
level $M=10$.}
\label{nsim10000}
\end{figure}

We first report the scatterplots of the posterior predictive
draws compared with i.i.d. draws from the target copula distributions based
on sample size $N=10,000$, as shown by Figure \ref{nsim10000}. The plots
come in pairs with the left one showing i.i.d. draws from the true copula
and the right one i.i.d. predictive draws from the posterior D-P tree to
compare. In most cases, our proposed D-P tree prior works well. The
difference between our predictive density and the true copula is mild, with
exceptions in highly correlated Gumbel case ($a=4$) and highly truncated
case (skew-normal, $\rho=0.9,\boldsymbol{\alpha}=(-10,50)$).

\begin{figure}[!h]
\centering
\FloatBarrier
\includegraphics[width=6.6in]{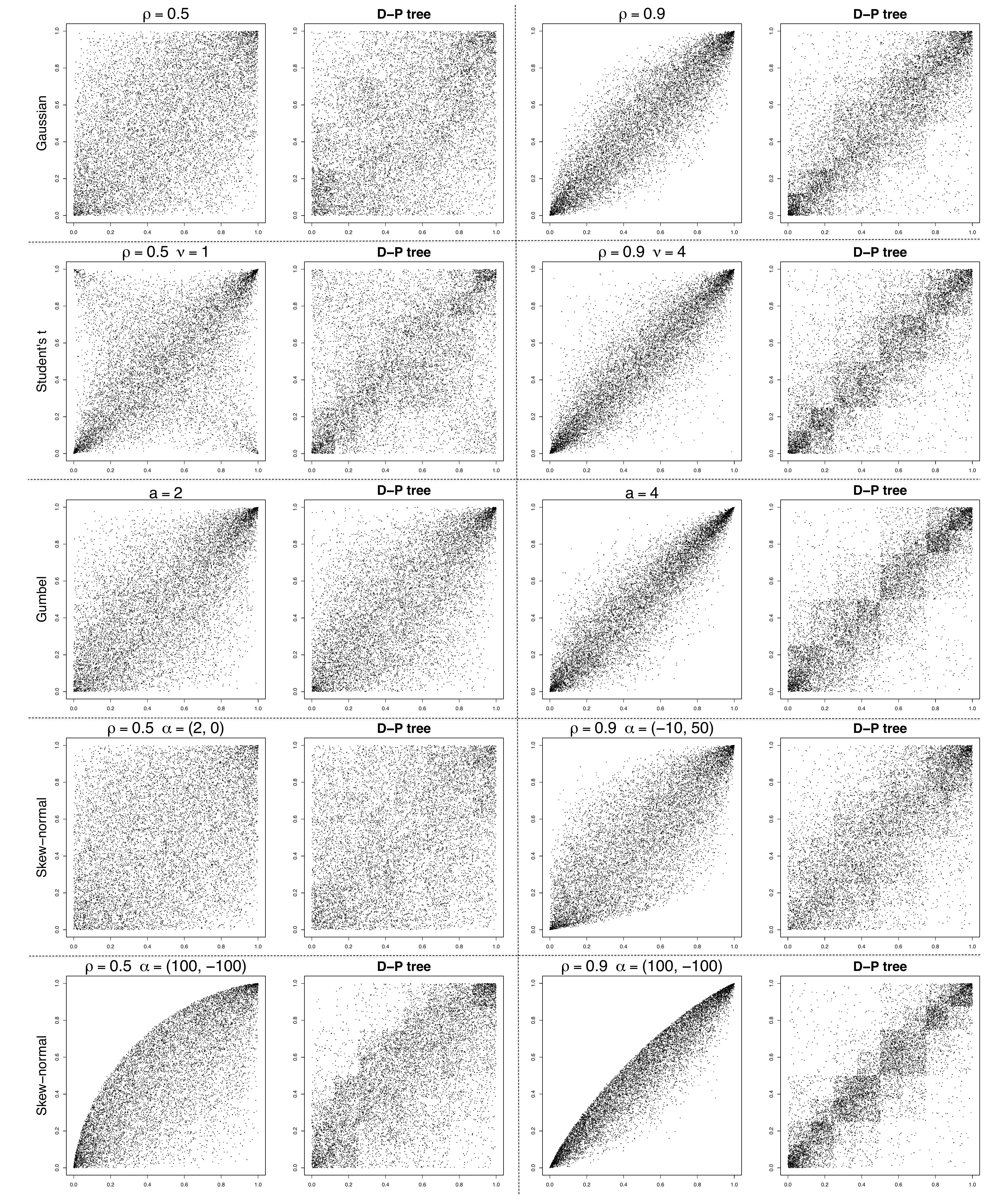}
\caption{Scatterplots of i.i.d. draws from the true copula distribution
(left) vs. the D-P tree posterior (right): sample size $N=1,000$, partition
level $M=10$.}
\label{nsim1000}
\end{figure}

\begin{figure}[!h]
\centering
\FloatBarrier
\includegraphics[width=6.6in]{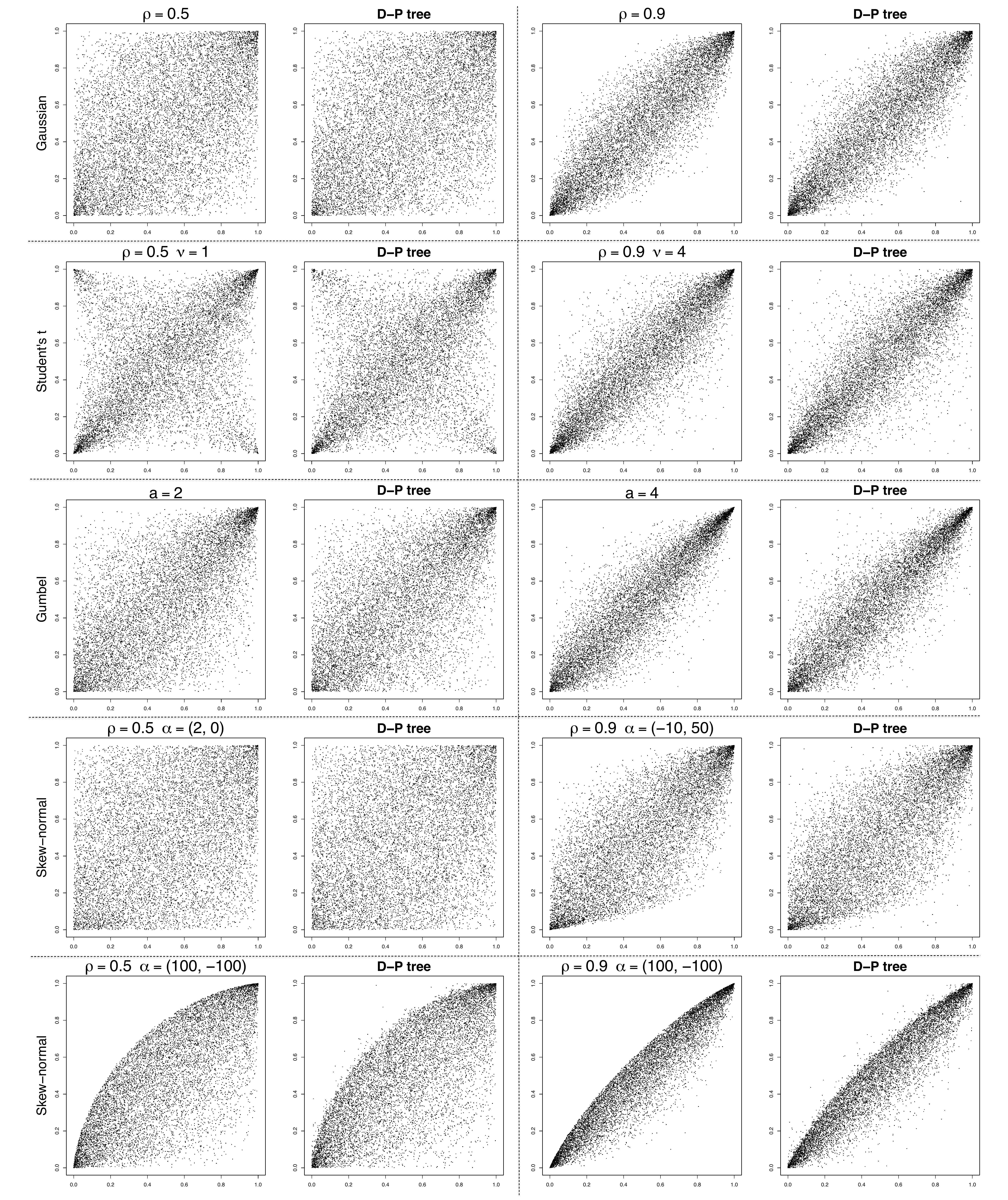}
\caption{Scatterplots of i.i.d. draws from the true copula distribution
(left) vs. the D-P tree posterior (right): sample size $N=100,000$,
partition level $M=10$.}
\label{nsim100000}
\end{figure}

In light of this, we increase the original sample size to $N=100,000$
(Figure \ref{nsim100000}). The overdispersion in low-density area and the
grid effect in highly correlated area are eliminated, which corresponds with
our asymptotic properties described in Section \ref{asymp}.

We also explored cases with the more challenging $N=1,000$. Here the
asymptotic conditions break, with $m=O(1)$ and $N/M^3=1$. Thereby, for more
complex copula structures, the D-P tree prior is prone to loss of
sensitivity due to the reduced sample size, which results in the rather
unfavorable grid effect as in Figure \ref{nsim1000}.

\subsubsection{Kullback-Leibler Divergence}

We further evaluate our method for copula density estimation quantitatively
using the Kullback-Leibler (K-L) divergence of our estimates from the true
copula:
\begin{align}
D_{KL}(C||\mathcal{P}_M)=\text{E}\left\{log(c/p_M)\right\}=\int\int\log\frac{%
c(u,v)}{p_M(u,v)}dC(u,v),
\end{align}
where $p_M$ is the density for $\mathcal{P}_M$, $c$ is the true copula
density, and the expectation is taken over the true copula distribution $C$ .

\begin{figure}[!h]
\centering
\includegraphics[width=3.5in]{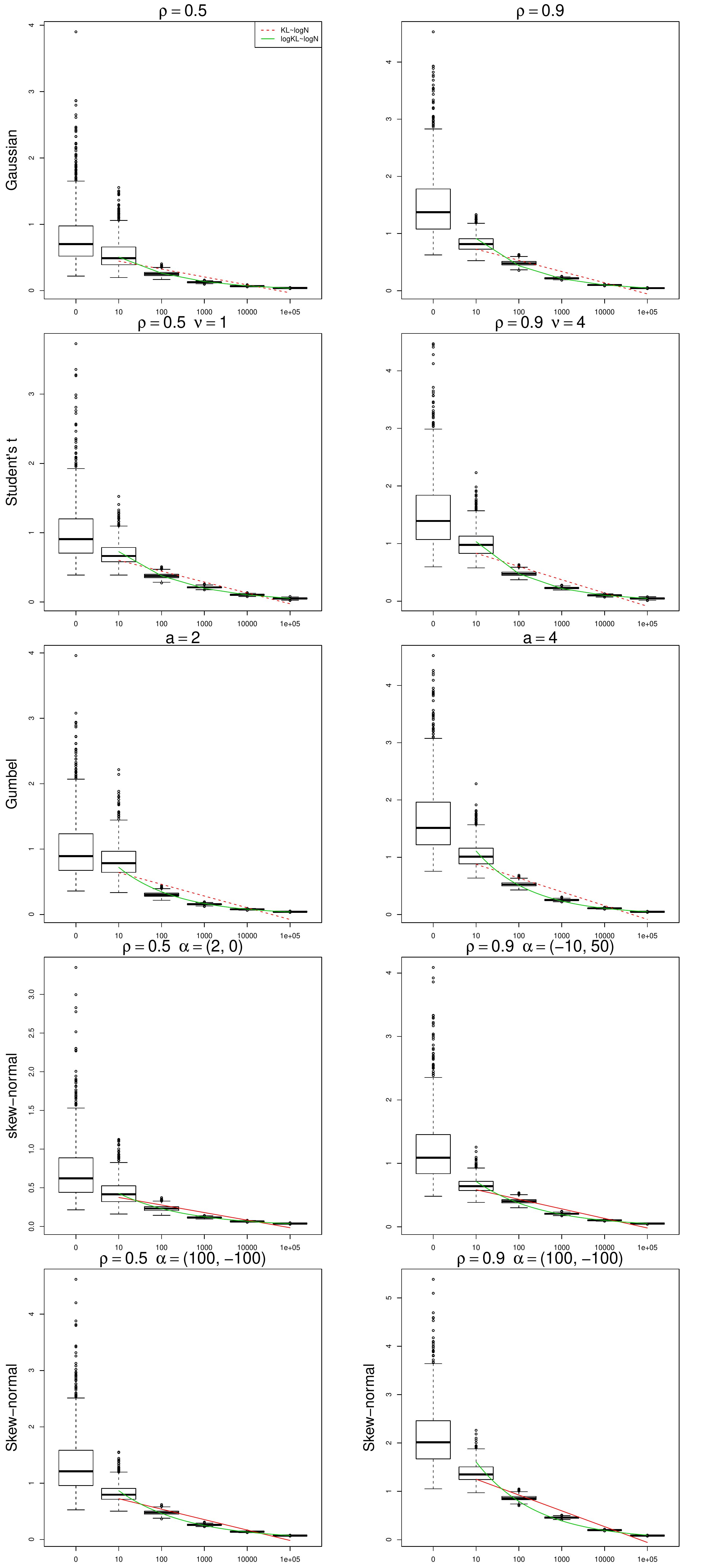}
\caption{Box-plots of the K-L divergence of the D-P tree posterior from the
target copulas against the sample size $N$: the solid green line showing a
linear fit of log(KL)$\sim$log($N$). }
\label{klplot}
\end{figure}

We vary the sample size $N$ from $0$ to $100,000$. For each sample size, we
draw 1,000 posterior densities from the D-P tree posterior and calculate the
K-L divergence using Monte-Carlo method. The mean and variance of the K-L
divergence are reported in Table \ref{klall} for various copula families, as
well as the box plots across various sample sizes by Figure \ref{klplot}.

Among all copula families, both mean and variance of the K-L divergence of the D-P tree posterior
converge to zero, showing the evidence of consistency in posterior.
Specifically the K-L divergence would drop below 0.10 when the sample size is
increased to 10,000, and the variance goes to 0, consistent with
the convergence claims in section \ref{asymp}.

To explore the convergence rate, we fit both $D_{KL}\sim \log N$ and $\log
D_{KL}\sim \log N $ with linear regression. The fitted curves are shown in
Figure \ref{klplot} by red and green respectively. The green curve gives
almost perfect fitting, indicating the convergence rate is in the order of $%
N^\alpha$, $\alpha<0$, though theoretical verification is still required.

%%%%K-L divergence
\begin{table}[!h]
\centering
\begin{tabular}{rr|llllll}
\hline\hline
\multicolumn{2}{c|}{$N$} & 0 & 10 & 100 & 1,000 & 10,000 & 100,000 \\ 
\hline\hline
\multicolumn{2}{r|}{$\rho$} & \multicolumn{6}{c}{Gaussian} \\ \hline
\multicolumn{2}{r|}{0.50} & 0.82{\scriptsize {(0.19)}} & 0.54{\scriptsize {%
(0.05)}} & 0.25 & 0.13 & 0.07 & 0.04 \\ 
\multicolumn{2}{r|}{0.90} & 1.50{\scriptsize {(0.35)}} & 0.83{\scriptsize {%
(0.02)}} & 0.48 & 0.22 & 0.10 & 0.05 \\ \hline\hline
$\rho$ & $\nu$ & \multicolumn{6}{c}{Student's t} \\ \hline
0.50 & 1.00 & 1.01{\scriptsize {(0.18)}} & 0.70{\scriptsize {(0.03)}} & 0.38
& 0.21 & 0.10 & 0.05 \\ 
0.90 & 1.00 & 0.86{\scriptsize {(0.21)}} & 0.60{\scriptsize {(0.05)}} & 0.24
& 0.14 & 0.07 & 0.04 \\ 
0.50 & 4.00 & 1.72{\scriptsize {(0.33)}} & 1.04{\scriptsize {(0.03)}} & 0.61
& 0.30 & 0.13 & 0.06 \\ 
0.90 & 4.00 & 1.53{\scriptsize {(0.37)}} & 1.01{\scriptsize {(0.05)}} & 0.48
& 0.23 & 0.10 & 0.05 \\ \hline\hline
\multicolumn{2}{r|}{$a$} & \multicolumn{6}{c}{Gumbel} \\ \hline
\multicolumn{2}{r|}{$2.00$} & 1.01{\scriptsize {(0.21)}} & 0.83{\scriptsize {%
(0.07)}} & 0.30 & 0.16 & 0.08 & 0.04 \\ 
\multicolumn{2}{r|}{$4.00$} & 1.66{\scriptsize {(0.39)}} & 1.04{\scriptsize {%
(0.05)}} & 0.53 & 0.25 & 0.11 & 0.05 \\ \hline\hline
{$\rho$} & {$\alpha$} & \multicolumn{6}{c}{skew-normal} \\ \hline
0.50 & (2,0) & 0.72{\scriptsize {(0.15)}} & 0.44{\scriptsize {(0.03)}} & 0.23
& 0.12 & 0.07 & 0.04 \\ 
0.50 & (-10,50) & 0.91{\scriptsize {(0.19)}} & 0.52{\scriptsize {(0.02)}} & 
0.31 & 0.18 & 0.09 & 0.05 \\ 
0.90 & (-10,50) & 1.22{\scriptsize {(0.27)}} & 0.65{\scriptsize {(0.01)}} & 
0.40 & 0.21 & 0.10 & 0.05 \\ 
0.90 & (50,0) & 1.09{\scriptsize {(0.22)}} & 0.65{\scriptsize {(0.04)}} & 
0.37 & 0.16 & 0.08 & 0.04 \\ 
0.50 & (100,-100) & 1.35{\scriptsize {(0.30)}} & 0.82{\scriptsize {(0.02)}}
& 0.48 & 0.26 & 0.14 & 0.07 \\ 
0.90 & (100,-100) & 2.13{\scriptsize {(0.39)}} & 1.39{\scriptsize {(0.04)}}
& 0.86 & 0.46 & 0.20 & 0.08 \\ \hline\hline
\end{tabular}%
\caption{Estimated K-L divergence of the D-P tree posterior from various targets, with
standard errors (SE). Note that we leave out SEs (all $0.00$) for $N\geq 100$%
. }\label{klall}
\end{table}

\subsection{Comparison with Existing Methods}

Here we compare our method with several existing nonparametric methods for copula
estimation.%, including, in the Bayesian settings: (a) the infinite Gaussian-mixture model, and in the frequentist settings: (2) the empirical copula estimation, (3) the histogram estimation, and (4) the kernel estimation methods.

\subsubsection{Comparison with Nonparametric Bayesian Methods}

We first compare our method with the infinite Gaussian mixture copula model %
\citep{wu2013bayesian}. For copula distribution $C$, we have the prior $%
C\sim\sum_{i=1}^{\infty}w_iC_g(\rho_i)$, where $C_g$ indicates the bivariate
Gaussian copula, and the weight $w_i\overset{i.i.d.}{\sim} U[0,1]$ and the
correlation $\rho_i\overset{i.i.d.}{\sim} U[-1,1]$. Such a model is the most
common one among existing nonparametric Bayesian methods which focus on
mixture models based on a specific copula family.

Here, we focus on the non-symmetric skew-normal copulas as the data generating copulas. The
simulations are carried out with the sample size varying from $N=1,000$ to $%
N=100,000$, and the K-L divergences of the estimates from the true target
copula distribution for both methods are calculated with Monte-Carlo method.
We report in Table \ref{klmixnorm} the cases where the skew-normal copula is
highly non-symmetric, and thus the Gaussian mixture model is severely misspecified.
More comprehensive simulation results on various target copulas are shown
in Supplementary Material \ref{append: sim}. For less non-symmetric copulas ($%
\alpha=(2,0),(-10,50),(50,0)$), the Gaussian mixture model dominates due to
its parametric nature, yet for these highly non-symmetric cases $%
(\alpha=(100,-100))$ in Table \ref{klmixnorm}, the D-P tree shows a
gradually increasing advantage as the data size increases. The results also
illustrate the inconsistency issue of the Gaussian mixture model, as its K-L
divergence from the data-generating model remains stable (0.17, 0.16) as
sample size increases, while the converging trend for the D-P tree posterior
is evident.

\begin{table}[!h]
\centering
\begin{tabular}{rr|ll|ll|ll}
\hline\hline
\multirow{3}{*}{$\rho$} & \multirow{ 3}{*}{$\alpha$} & \multicolumn{6}{c}{$N$%
} \\ \cline{3-8}
&  & \multicolumn{2}{c}{1,000} & \multicolumn{2}{c}{10,000} & 
\multicolumn{2}{c}{100,000} \\ \cline{3-8}
&  & D-P & GM & D-P & GM & D-P & GM \\ \hline
0.50 & (100,-100) & 0.26 & 0.17 & 0.14 & 0.17 & 0.07 & 0.17 \\ 
0.90 & (100,-100) & 0.46 & 0.16 & 0.20 & 0.16 & 0.08 & 0.16 \\ \hline\hline
\end{tabular}%
\caption{Comparison of the K-L divergence between the D-P tree (D-P) and the Gaussian mixture (GM) model for highly
non-symmetric skew-normal target copulas.}\label{klmixnorm}
\end{table}

%%%%%simulation 

\subsubsection{Comparison with Nonparametric Frequentist Methods}

We select three classic nonparametric methods in frequentist settings in
comparison with our D-P tree. Suppose $Y_{i}=(U_{i},V_{i})\overset{i.i.d.}{\sim }C$:

\begin{itemize}
\item \textit{The empirical estimator:} $\hat{C}_{emp}(u,v)=\frac{1}{N}\sum_{i=1}^{N}I_{U_{i}\leq u}I_{V_{i}\leq v}$.

\item \textit{The histogram estimator:} $\hat{C}%
_{hist}(B_{\epsilon })=\frac{n_{\epsilon }}{N}$, where $B_{\epsilon }$ is
the partition at the highest level.

\item \textit{The independent Gaussian kernel estimator \citep%
{jaworski2010copula}:}  
\begin{align}
\hat C_{ker}(u,v)=\frac{1}{N}\sum_{i=1}^{N} \Phi\left\{\frac{%
\Phi^{-1}(u)-\Phi^{-1}(U_i)}{h}\right\}\Phi\left\{\frac{\Phi^{-1}(v)-%
\Phi^{-1}(V_i)}{h}\right\},
\end{align}
where we make the choice of $h=N^{-\frac{1}{5}}$, following Silverman's
rule of thumb for the choice of window width.

\item \textit{The D-P tree posterior mean estimator:} for a
fair comparison, we use the mean distribution from the D-P tree posterior as
the Bayesian estimator by the D-P tree, i.e., $\hat C_{D-P}=\text{E}%
(C|\boldsymbol{Y})$.
\end{itemize}

We define several measurements for the distance between the estimator and
the target distribution. For density estimation, besides the K-L divergence,
we also include the commonly adopted MISE (Mean Integrated Squared Error) based on the averaged $%
L_2 $-norm between the estimated density function and the truth: 
\begin{align}
MISE(\hat c)=\text{E}\left[\iint_{[0,1]\times[0,1]}\{c(u,v)-\hat
c(u,v)\}^2du\,dv\right].
\end{align}
Here $c$ is the target copula density and $\hat c$ is its estimator.

For the distance measurement of the distribution, we extend the $MISE$ for
density to the $MISE_C$ (Mean Integrated Squared Error for Cumulative
functions): 
\begin{align}
MISE_C(\hat C)=\text{E}\left[\iint_{[0,1]\times[0,1]}\{C(u,v)-\hat
C(u,v)\}^2du\,dv\right],
\end{align}
where $\hat C$ is the estimated copula function, $C$ is the
true copula. We also have a distance measure specifically targeting the
grid-based estimation methods, the $MSE_g$: 
\begin{align*}
MSE_g(\hat C)=\text{E}\left[\frac{1}{2^{2M}}\sum_{i,j=1}^{2^M}\{C(B_{ij})-%
\hat C(B_{ij})\}^2\right],
\end{align*}
where $\{B_{ij}\}$ are partitions on $[0,1]\times[0,1]$, and $M$ is the maximum
partition level. Note that all the expectations in the measures defined above
are taken over all possible data samples

The simulations are carried out with the sample size varying from $N=10$ to $%
N=10,000$ for a good look at the convergence trend. We again focus mainly on
heavily non-symmetric skew-normal copulas ($\alpha=(-50,10), (100,-100)$).
For each parameter setting, we first draw $N$ i.i.d. samples from the true
copula distribution, obtain the copula estimates by three frequentist
methods and the D-P tree posterior mean estimator; then we repeat this
process 50 times to obtain the Monte-Carlo approximation of the measures as defined above. Note that for the empirical
copula estimation, the estimated distribution is discrete, thus the
density distance measures not applicable; for the histogram estimator, due
to the discrepancy in the supports between the target and the estimated
distributions, the K-L divergence is not applicable. To ensure computational
efficiency, we report the results based on the approximation level $M=8$,
and to maintain comparability, we take the same maximum partition level for the histogram estimation method. Comprehensive numeric results are
shown in Table \ref{table: kl all}-\ref{table: mseg all} in Supplementary Material \ref%
{append: sim}. Here we report mainly the results under the parameter setting 
$\rho=0.5, \alpha=(100,-100)$ in Table \ref{table: freq} as exemplary for
our conclusions.

\begin{table}[!h]
\centering
\begin{tabular}{r|cccc|cccc}
\hline\hline
\multirow{2}{*}{$N$} & \multicolumn{4}{c|}{K-L} & \multicolumn{4}{c}{$\sqrt{%
MISE}$} \\ \cline{2-9}
& D-P Tree & Empirical & Kernel & Hist. & D-P Tree & Empirical & Kernel & 
Hist. \\ \hline
10 & 0.528 & NA & 0.528 & Inf & 1.365 & NA & 2.190 & 71.788 \\ 
20 & 0.473 & NA & 0.428 & Inf & 1.163 & NA & 2.657 & 56.726 \\ 
50 & 0.386 & NA & 0.314 & Inf & 1.050 & NA & 1.177 & 36.757 \\ 
100 & 0.349 & NA & 0.261 & Inf & 1.159 & NA & 1.347 & 25.723 \\ 
500 & 0.222 & NA & 0.166 & Inf & 1.072 & NA & 1.398 & 11.665 \\ 
1,000 & 0.184 & NA & 0.136 & Inf & 0.894 & NA & 0.703 & 8.078 \\ 
5,000 & 0.112 & NA & 0.090 & Inf & 0.703 & NA & 0.516 & 3.601 \\ 
10,000 & 0.089 & NA & 0.076 & Inf & 0.701 & NA & 0.769 & 2.600 \\ 
\hline\hline
& \multicolumn{4}{c|}{$\sqrt{MISE_C}$} & \multicolumn{4}{c}{$\sqrt{MSE_g}$}
\\ \hline
10 & 0.072 & 0.118 & 0.091 & 0.117 & 0.054 & 0.321 & 0.054 & 0.321 \\ 
20 & 0.065 & 0.082 & 0.068 & 0.082 & 0.054 & 0.230 & 0.055 & 0.230 \\ 
50 & 0.044 & 0.057 & 0.050 & 0.057 & 0.054 & 0.151 & 0.054 & 0.151 \\ 
100 & 0.037 & 0.041 & 0.038 & 0.041 & 0.054 & 0.113 & 0.054 & 0.113 \\ 
500 & 0.018 & 0.017 & 0.017 & 0.017 & 0.053 & 0.070 & 0.053 & 0.070 \\ 
1,000 & 0.013 & 0.012 & 0.013 & 0.012 & 0.053 & 0.062 & 0.053 & 0.062 \\ 
5,000 & 0.007 & 0.006 & 0.007 & 0.006 & 0.053 & 0.055 & 0.053 & 0.055 \\ 
10,000 & 0.005 & 0.004 & 0.005 & 0.004 & 0.053 & 0.054 & 0.053 & 0.054 \\ 
\hline\hline
\end{tabular}%
\caption{Comparison of various distance measures between the D-P tree
posterior mean estimator and frequentist estimators for the skew-normal copula with parameter $\protect\rho=0.5$, $\protect\alpha=(100,-100)$%
. }\label{table: freq}
\end{table}

%%%%%%
In general, the D-P tree posterior mean estimator performs competitively
well compared with all three frequentist nonparametric methods and
consistently across various measures. Notably, the D-P tree
posterior estimation appears advantageous over other methods with small sets
of observations, showcasing a preferably strong smoothing effect
induced by the D-P tree prior.

Both the D-P tree and the kernel estimation show drastic advantages in
copula density estimation over empirical and histogram methods, as the
empirical copula fails to yield density estimator and the histogram
estimator gives severely poor density approximation due to the discrepancy
in the support. Though both methods take advantage of the smoothing effect
in estimation density, under the MISE measurement, the D-P tree dominates
kernel method across almost all sample sizes while giving close figures
under the K-L divergence.

As for copula distribution estimation, the D-P tree shows a strong advantage
over other methods in both measures under scenarios of smaller sample
size, which indicates the more favorable continuity feature of
the D-P tree prior. When the sample size increases, the neutralizing effect
of the D-P tree prior slows down the convergence of the posterior, and thereby,
the empirical and histogram estimators catch up in figures. Yet still, up to 
$N=10,000$, the D-P tree gives close distances as the empirical and the
histogram methods, and consistently dominates the kernel method.

%%%%%application

\section{Real data application}

\label{sec: app} For real data analysis, we apply our method to the S\&P 500
daily index and the IBM daily stock prices over the past 20 years (Jan 1,
1994 to Dec 31, 2014, available from \url{https://finance.yahoo.com}) and aim to estimate their dependence structure with
the copula model. We adopt both the cross-validation and the rolling
prediction schemes to evaluate the performance of our method, as described
below in detail. We assess three methods under both prediction schemes for
comparison: (1) the D-P tree posterior mean with canonical prior; (2) the
Gaussian copula; (3) the Student's t copula. For the rolling prediction
assessment, we also include (4) the independent Gaussian kernel estimator;
(5) the empirical copula estimator, and (6) the D-P tree posterior mean with
historic-data-induced prior. Typically, investment groups have focused on
using methods (2), (3) and particularly (5) in practice for risk management.

\subsection{Cross-validation}

We first conduct cross-validation to evaluate the prediction ability of our
method. Let the joint daily prices for two stocks be $%
\{(y_{i}^{1},y_{i}^{2}),\,i=1,\dots ,T\}$, where $T=5,288$, and the returns
of log price $\{r_{i}^{j}=\log y_{i}^{j}-\log y_{i-1}^{j},\,i=2\dots
T,\,j=1,2\}$. Marginally, we fit the commonly adopted GARCH(1,1) model: $%
r_{i}^{j}=\sigma _{i}^{j}\epsilon _{i}^{j}$, $(\sigma _{i}^{j})^{2}=\alpha
_{0}^{j}+\alpha _{1}^{j}(\sigma _{i-1}^{j})^{2}+\beta _{1}^{j}(\epsilon
_{i-1}^{j})^{2}$, where the innovations $\{\epsilon _{i}^{j}\}$ are
independent with $\text{E}(\epsilon _{i}^{j})=0$ and $\text{var}(\epsilon
_{i}^{j})=1$. Further, we assume the distribution of the innovations is
time-invariant and put the copula model on their joint distribution $%
F(\epsilon _{i}^{1},\epsilon _{i}^{2})=C(F^{1}(\epsilon
_{i}^{1}),F^{2}(\epsilon _{j}^{2}))$, where $(\epsilon _{i}^{1},\epsilon
_{i}^{2})\overset{i.i.d.}{\sim }F$, and $F^{1}$ and $F^{2}$ are the marginal
distributions.

\begin{figure}[h]
\centering
\includegraphics[width=5in]{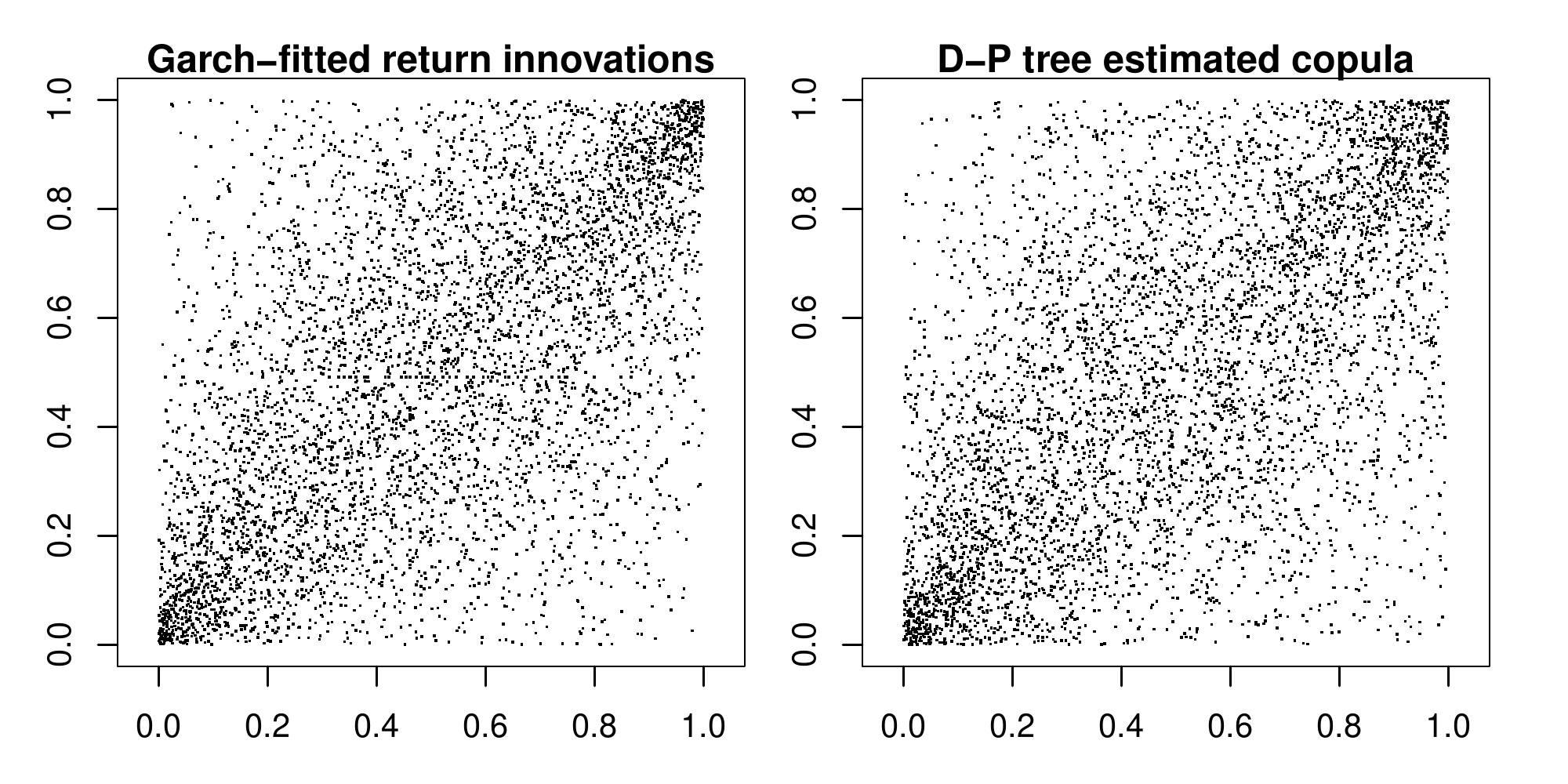}
\caption{Scatterplots comparing the GARCH-fitted joint return innovations
(left, normalized by the empirical marginal distributions) and random draws
from the D-P tree estimated copula (right).}
\label{fig:stock}
\end{figure}

We apply the proposed D-P tree (canonical) prior to the copula estimation
based on the fitted innovations from the GARCH model $\{(\hat\epsilon^1_i,%
\hat\epsilon^2_i)\}$. Here we use the empirical estimation for the
marginals, as discussed in Section \ref{sec:margin}. Figure \ref{fig:stock}
compares the scatterplots of the fitted errors (normalized by marginals) and
the draws from the D-P tree posterior. We observe no apparent discrepancy
between the data and the fitted model.

For comparison, we also fit the innovations with the bivariate Gaussian and
Student's t copula models respectively, as described in Appendix \ref%
{append:copula}. The estimates are obtained by maximum likelihood estimation.

To assess the effectiveness of such estimation methods, we adopt the
cross-validation scheme to get the estimation of the prediction errors. We
randomly divide the marginally GARCH-fitted innovations into 10 sets and
each time use one set as the testing set and the rest as the training set.
For prediction errors, we adopt the Monte-Carlo estimation for cross entropy 
$-E_c(log(\hat c))$, which is equivalent to the K-L divergence of the
estimated copula density from the assumed truth up to a constant. Here, $c$
and $\hat c$ denote the true copula density and the estimated one
respectively.

Table \ref{tab:stock} gives the mean K-L divergence based on 10-fold
cross-validation for the three methods in comparison. As it shows, the D-P
tree prior outperforms the Gaussian copula narrowly, while the Student's t
shows an advantage over the other two methods. The result is unsurprising as
the distribution of stock returns are notable for the heavy-tail dependence
features and the Student's t is thus expected to give good fittings.
Nevertheless, when parametric models are misspecified under the Gaussian
copula model, nonparametric methods such as the D-P tree prior is still
advantageous.

\begin{table}[!h]
\centering
\begin{tabular}{r|rrr}
\hline\hline
& D-P tree & Gaussian ($\hat\rho$=0.59) & Student's t ($\hat\rho$=0.60, $%
\hat\nu$=6.5) \\ \hline
Cross entropy & -0.210 & -0.209 & -0.224 \\ \hline\hline
\end{tabular}%
\caption{Comparison of the mean prediction errors based on the K-L
divergence between the D-P tree prior, the Gaussian copula and the Student's
t copula. Note that the more negative the numbers, the better the prediction
performance.}\label{tab:stock}
\end{table}

\subsection{Rolling Prediction}

To mimic the practical prediction scenario, we also evaluate the prediction
power of our method under the time-rolling prediction scheme, that is, we
predict the future copula structure within a certain window of time based on
the most recent observations.

Specifically, we set a training length of $T_{tr}$, a testing set length of $%
T_{te}$, a rolling estimation window of length $t_{e}$, and a prediction
window of length $t_{p}$. Firstly, we use the daily price time series of the
two stocks $\{y_{t}^{1}:t=1,\dots ,{T_{tr}}\}$ and $\{y_{t}^{2},t=1,\dots ,{%
T_{tr}}\}$ as the training set for the marginal GARCH-model fitting.
Consistent with common practical prediction scenarios, we fix such fitted
GARCH model and obtain the fitted innovations for the training set $\{(\hat{%
\epsilon}_{t}^{1},\hat{\epsilon}_{t}^{2}),t=1,\dots ,{T_{tr}}\}$, and the
predicted innovations for the test set $\{(\hat{\epsilon}_{t}^{1},\hat{%
\epsilon}_{t}^{2}),t=T_{tr}+1,\dots ,{T_{tr}+T_{te}}\}$. Then, we conduct
the rolling prediction of the copula structure based on these estimates. For
each rolling step, we apply the proposed D-P tree-based method with both the
canonical non-informative prior and the historic-data-induced prior to the
most recent $t_{e}$-fitted/predicted innovations and estimate the future
copula structure of length $t_{p}$. Here we implicitly assume the i.i.d.
property of the innovations within the estimation and prediction windows
combined of length ($t_{e}+t_{p}$). This is reasonable in that the copula
structure is usually stable within a certain length of time. We repeat such
rolling prediction $T_{te}/t_{p}$ times until the whole testing length ($%
T_{tr}+1$ to $T_{tr}+T_{te}$) is covered.

Here we focus on the data of the period covering the 2007-08 financial
crisis (i.e., the testing set covering July, 2007 to July, 2009) to
highlight the flexibility and robustness of nonparametric methods over
traditional parametric models. We set $T_{tr}=500$, $T_{te}=500$, and vary $%
t_e\in\{10,20,50,100,250\}$, $t_p\in\{1,50\}$ and report both the average
log-likelihood $\frac{1}{T_{te}}\sum_{t=1}^{T_{te}}\log\hat{c_t}$
(equivalent to negative KL divergence plus a constant), and the square root
of average ${MISE_C}=\frac{1}{T_{te}}\sum_{t=1}^{T_{te}}MISE_C(\hat{C_t})$
as the measures for prediction accuracy (Table \ref{tab:stock.roll}). Note
that for historic-data-based D-P tree prior, we adopt the posterior of a
canonical D-P tree prior updated by the data from testing set ($%
i=1,\dots,T_{tr}-t_e$) with each down-weighted by 0.1. We also carry out the
same prediction scheme with other four methods for comparison.

\begin{table}[ht]
\scriptsize
\centering
\begin{tabular}{rr|llllll|llllll}
\hline\hline
\multirow{2}{*}{ $t_{e}$} & \multirow{2}{*}{ $t_{p}$} & \multicolumn{6}{c}{
Average log-likelihood} & \multicolumn{6}{c}{$\sqrt{MISE_C}$} \\
 \cline{3-14} 
&  & D-PT & D-PTw & Emp. & Kernel & Gauss. & t & D-PT & D-PTw & Emp. & Kernel & Gauss. & t \\ \hline
10 & 1 & -0.002 & \bf0.133 & NA & -0.052 & 0.094 & 0.086 & 0.312 & \bf0.300 & 0.338 & 0.305 & \bf0.300 & 0.301 \\ 
20 & 1 & 0.046 & 0.135 & NA & 0.030 & \bf0.141 & 0.139 & 0.310 & 0.301 & 0.328 & 0.305 & \bf0.299 & 0.300 \\ 
50 & 1 & 0.096 & 0.141 & NA & 0.044 & 0.141 & \bf0.143 & 0.310 & 0.304 & 0.322 & 0.306 & \bf0.299 & \bf0.299  \\ 
100 & 1 & 0.155 & \bf0.176 & NA & 0.096 & 0.154 & 0.160 & 0.309 & 0.306 & 0.318 & 0.306 & \bf0.298 & 0.299 \\ 
250 & 1 & 0.173 & \bf0.178 & NA & 0.105 & 0.153 & 0.158 & 0.306 & 0.306 & 0.312 & 0.304 & \bf0.298 & \bf0.298 \\ 
10 & 50 & 0.023 &\bf 0.138 & NA & -0.075 & -0.340 & -0.128 & 0.082 & \bf0.062 & 0.123 & 0.099 & 0.078 & 0.074  \\ 
20 & 50 & 0.051 &\bf 0.137 & NA & 0.003 & 0.009 & 0.028 & 0.082 & \bf0.064 & 0.102 & 0.091 & 0.071 & 0.071 \\ 
50 & 50 & 0.113 & \bf0.156 & NA & 0.058 & 0.106 & 0.113 & 0.066 & \bf0.060 & 0.070 & 0.068 & 0.066 & 0.066\\ 
100 & 50 & 0.155 &\bf 0.173 & NA & 0.067 & 0.137 & 0.139 & 0.060 & \bf0.058 & 0.063 & 0.062 & 0.063 & 0.063 \\ 
250 & 50 & 0.175 & \bf0.181 & NA & 0.108 & 0.150 & 0.158 & \bf0.057 & \bf0.057 & 0.058 & 0.058 & 0.061 & 0.061\\ 
\hline\hline
\end{tabular}%
\caption{Comparison of the prediction performance in the average
log-likelihood (the higher the numbers, the better the prediction) and the $%
MISE_C$ (the lower, the better) between various methods: the D-P tree
posterior mean with the canonical prior (D-PT), the D-P tree with the
historic-data-induced prior (D-PTw), the empirical copula (Emp.), the kernel
estimator (Kernel), the Gaussian copula (Gauss.) and the Student's t
copula (t) models.}\label{tab:stock.roll}
\end{table}

Generally, both the D-P-tree-based methods show strong advantages over other
methods by the log-likelihood loss in almost all settings, and by $\sqrt{%
MISE_C}$ under a longer prediction window $t_p=50$ (where the
distribution-based measure $\sqrt{MISE_C}$ is more valid due to multiple testing samples) and a larger
prediction set $t_e\geq 50$. Such results verify the robustness and
adaptiveness of the D-P tree-based methods to irregular market behaviors
when classic parametric models are terribly misspecified. Further, by
incorporating the historic data into the prior, the D-PTw method enjoys a
strong boost in prediction accuracy, and dominates other methods in most of
the scenarios. Admittedly, more data are used by the D-PTw for inference
than other methods in comparison. Nevertheless, it is exactly the showcase of the
strength of Bayesian methods where historic or empirical information is
readily concocted into priors to help.

\section{Discussion}

\label{sec: diss}

\subsection{Copula Normalizing}

One problem with most nonparametric copula estimation methods including the
D-P tree prior is that the posterior marginal does not always follow a
uniform distribution. Suppose $\mathcal{P}\sim DPT(\Pi,\mathcal{A}|%
\boldsymbol{Y})$, then marginally $\mathcal{P}([0,1/2]\times[0,1])\sim
Beta(\alpha_0+n_0+\alpha_1+n_1,\alpha_2+n_2+\alpha_3+n_3)$, which deviates
from $0.5$ by the randomness. Though, when the sample size $N$ is large, as
shown by Proposition \ref{prop: pw convg}, the posterior density would have
marginals close to uniforms, thus approximate a proper copula density, the
issue of normalizing posterior density to proper copula density still needs
addressing. Here we provide several methods to carry out the 
correction.

\subsubsection{Ad Hoc Correction}

Suppose we have $\mathcal{P}^{\ast }\sim DPT(\Pi ,\mathcal{A}|Y$) and $%
\mathcal{P}_{M}^{\ast }$ is its M-level approximation with a $2^{M}\times
2^{M}$ grid density. To normalize its marginals to the uniforms, we need to
restrain 
\begin{equation}\label{gridprob}
\mathcal{P}_{M}^{\ast }([k/2^{M},(k+1)/2^{M}]\times \lbrack 0,1])=\mathcal{P}%
_{M}^{\ast }([0,1]\times \lbrack k/2^{M},(k+1)/2^{M}])=1/2^{M}, 
\end{equation}%
$k=0,1,\dots ,2^{M}-1$; i.e., the column sum and row sum of the $2^{M}\times
2^{M}$ grid density to be $1/2^{M}$. One way to realize this is to randomly
select $2\cdot 2^{M}-1$ grids and manipulate
their values to fit \eqref{gridprob}.

As $\mathcal{P}^*_M$ is close to $C$ when the sample size is large, the
marginals of $\mathcal{P}_m$ would not be too far away from a uniform. Thus
the ad hoc correction would not cause severe deviation from the posterior
density $\mathcal{P}^*$.

\subsubsection{Inverse Transform on the Marginals}

Another way of normalization is to apply the PIT (Probability Inverse
Transform) to the marginals of $\mathcal{P}^*_M$. Factorize the M-level
approximate posterior density by 
$
\mathcal{P}_M^*([0,x]\times[0,y])=C_{\mathcal{P}_M^*}(F_{x,\mathcal{P}%
_M^*}(x),F_{y,\mathcal{P}_M^*}(y)),
$
where $F_{x,\mathcal{P}_M^*}$ and $F_{y,\mathcal{P}_M^*}$ are the marginal
CDFs of $\mathcal{P}_M^*$, and $C_{\mathcal{P}_M^*}$ is their copula. By
transforming $(x,y)\to(F_{x,\mathcal{P}_M^*}(x),F_{y,\mathcal{P}%
_M^*}(y))=(u,v)$, we have the normalized distribution $\mathcal{\tilde{P}}%
_M^*$: 
\begin{align*}
\mathcal{\tilde{P}}_M^*([0,u]\times[0,v])=C_{\mathcal{P}_M^*}(u,v)=\mathcal{P%
}_M^*([0,F^{-1}_{x,\mathcal{P}_M^*}(u)]\times[0,F^{-1}_{y,\mathcal{P}_M^*}(v)%
]),
\end{align*}
which is a proper copula distribution.

One good property of such normalization is that it preserves the copula
structure due to the monotonicity of the transform, i.e., $\mathcal{P}^*_m$
and $\mathcal{\tilde{P}}_m^*$ share the same copula. Further,
asymptotically, $F_{x,\mathcal{P}_m^*}$ and $F_{y,\mathcal{P}_m^*}$ converge
to the uniforms, leading to $\mathcal{\tilde{P}}_m^*\overset{p}{\to}\mathcal{%
P}^*_m$.

\subsection{Estimation with Unknown Marginals}

\label{sec:margin} Throughout this article, especially for the simulations,
we focus on the estimation of a copula itself, assuming the marginals are
known. Here we address more practical scenarios where the marginals are to
be estimated. As we stated earlier, the marginal distributions can be more
accurately estimated than the copula as data concentrate to a single
dimension. Generally, suppose we have $N$ i.i.d. observations $(X_i, Y_i)$,
and their marginal distribution estimates are either parametric or
nonparametric, denoted by $\hat F_X$ and $\hat F_Y$ respectively. The
inverse transform $(\hat F_X^{-1}(X_i), \hat F_Y^{-1}(Y_i))=(\hat U_i, \hat
V_i) $ is considered copula-distributed observations where the regular D-P
tree copula estimation procedure can be applied. 

\iffalse
Specifically, with a parametric model on the
marginals as $F_\psi$, we could adopt a semi-parametric framework
combining the parametric marginal estimation, supposedly $F_{X,\hat\psi}$, $%
F_{Y,\hat\psi}$, and the nonparametric D-P tree copula estimation based on
the inverse transformed data $(F_{X,\hat\psi}^{-1}(X_i),
F_{Y,\hat\psi}^{-1}(Y_i))$. Yet, as the copula is
increasing-monotone-transform invariant, a more coherent approach is to
adopt a full nonparametric framework where the marginals are first
estimated empirically and then approximately transformed to uniforms by the
empirical quantiles for the next-step copula estimation. Within such a full
nonparametric method, only the rank information is preserved, which is
sufficient for nonparametric copula estimation.
\fi

\subsection{Higher Dimension}
Most of the results of the D-P tree prior on bivariate copulas can be generalized to higher dimensions.
Specifically, for a $d$-dimensional copula, we can generalize the D-P tree
prior to $C\sim DPT(\Pi, \mathcal{A})$, where $\Pi$ is a $2^d$-partition on
the $d$-dimensional unit cube and the same parametrization for $\mathcal{A}%
=\{\alpha: \alpha_{\epsilon_1\dots\epsilon_m}=m^2\}$. Those properties of a
bivariate D-P tree including conjugacy, continuity and convergence, are
still preserved.

However, as the dimension increases, the sparsity of data would cause great
difficulty for accurate copula estimation, especially among nonparametric
settings including the D-P tree prior. Further,
though the computational complexity is stable, the D-P tree still requires
exponentially increasing storage power as the dimension increases. Yet one
potentially favorable feature of the D-P tree that we have observed through
simulations is its strong smoothing effect and improved estimation accuracy when the sample size is small. Thereby, the
D-P tree prior could be the more favorable nonparametric method compared to
other alternatives with sparse observations under higher-dimensional
scenarios. This could be one potential angle for further studies.

\section{Conclusion}

\label{sec: concld} The proposed Dirichlet-based P\'olya tree (D-P tree)
prior preserves properties including conjugacy, continuity and convergence
as the classic P\'olya tree, which provides a foundation for nonparametric
copula estimation under the Bayesian framework. Compared with other Bayesian
copula estimation methods, the D-P tree prior exhibits strength in
robustness and consistency, remedying the severe bias of the earlier P\'olya
tree-based prior in copula estimation, and also overcoming the inconsistency
issue of the family-based mixture model under misspecification. In
comparison with the nonparametric methods under the frequentist settings,
the D-P tree posterior mean estimator performs competitively well and rather
stably across various distance measures. Notably, with a small sample size,
the D-P tree copula estimator is advantageous in estimation accuracy, which
may imply its potential in higher-dimensional cases where observations are
heavily diluted.

However, there still are issues remaining with the D-P tree prior worthy of
further exploration, such as the marginal bias caused by the randomness in
the prior and a more efficient application in higher dimensions. Further, in
terms of the D-P tree's application to the copula prediction of the stock
prices, we have not yet fully exploited the timely nature of the data. The
exploration of time-dependent D-P tree prior could be of great future
research interest. In addition, alternative priors under nonparametric
Bayesian frameworks could also be of future interest to overcome the
limitations of the D-P tree prior.

%TCIMACRO{\TeXButton{Appendix}{\appendix}}%
%BeginExpansion
\appendix%
%EndExpansion

%\appendix

\begin{center}
\bf \LARGE
\uppercase {Appendix}
\end{center}

\section[Appendix]{Definitions}

\subsection{Copula}
\begin{mydef}
$C: [0,1]^d \to [0,1] $ is a d-dimensional copula, if $C$ is a joint
cumulative distribution function for a d-dimensional random vector on $%
[0,1]^d$ with uniform marginals. For two-dimensional case, that is, $%
C(u,v)=P(U\leq u,V\leq v)$, where $U,V\sim Unif[0,1]$. And the joint density
function $c(u,v)$ is called copula density.
\end{mydef}

\textbf{Sklar's Theorem} \citep{sklar1959fonctions}, if $X$ and $Y$ are
random variables with cumulative distribution functions $F$ and $G$, and a
joint distribution function $H$, then there exists a copula $C$ such that
for all $(x,y)\in \mathbb{R}^{2}$, 
$
H(x,y)=C(F(x),G(y)),
$
and for density function, we have $h(x,y)=c(F(x),G(y))f(x)g(y)$, where $f$
and $g$ are marginal density functions and $h$ is the joint density.
\iffalse
When both $F$ and $G$ are continuous, copula $C$ is unique and $X%
\rotatebox[origin=c]{90}{$\models$} Y$ is equivalent to their copula being $%
C(u,v)=uv$.

\textbf{Fr\'{e}chet-Hoeffding Theorem: }$W(u,v)\leq C(u,v)\leq M(u,v)$,
where $W(u,v)=\max \{u+v-1,0\}$, and $M(u,v)=\min \{u,v\}$.
\fi
\subsection{P\'{o}lya Tree}

\label{append:def copula}

\begin{mydef}
\citep{lavine1992some} Let $\Omega$ be a separable measurable space and $%
\Pi=\{B_\epsilon\}$ be one of its binary tree partitions that generate the
measurable sets, where $B_\emptyset=\Omega$ and $B_\epsilon=B_{\epsilon0}%
\cup B_{\epsilon1}$. A random probability measure $\mathcal{P}$ is said to
have a P\'olya tree distribution, or P\'olya tree prior, with parameters ($%
\Pi$,$\mathcal{A}$), written $\mathcal{P}\sim PT(\Pi,\mathcal{A})$ , if
there exists non-negative numbers $\mathcal{A}=\{\alpha_0,\alpha_1,%
\alpha_{00},\dots\}$ and random variables $\mathcal{Z}=\{Z_0,Z_1,Z_{00}\dots%
\}$ such that the following hold:

\begin{itemize}
\item all the random variables in $\mathcal{Z}$ are independent;

\item for every $m=1,2,\dots$ and every $\epsilon=\epsilon_1\epsilon_2\dots%
\epsilon_m$, $Z_{\epsilon}\sim Beta(\alpha_{\epsilon0},\alpha_{\epsilon1})$;

\item for every $\epsilon$, 
$
\mathcal{P}(B_{\epsilon=\epsilon_1\epsilon_2\dots\epsilon_m})=\left(%
\prod_{j=1;\epsilon_j=0}^mZ_{\epsilon_1\epsilon_2\dots\epsilon_{j-1}}\right)%
\left\{\prod_{j=1;\epsilon_j=1}^m(1-Z_{\epsilon_1\epsilon_2\dots%
\epsilon_{j-1}})\right\}
$
where the first terms in the products are interpreted as $Z_{\emptyset}\sim
Beta(\alpha_0,\alpha_1)$ and $(1-Z_{\emptyset})$.
\end{itemize}
\end{mydef}

\section{Proofs}

\label{append:prf}

\subsection{Proof of Proposition \protect\ref{prop: pw convg}}

\label{prf: cvg} We consider $\mathcal{P_{M}}$ on the measurable partition $%
\{B_{\epsilon }\}$. For any $B_{k}=B_{\epsilon _{1}\dots \epsilon _{k}}\in
\Pi $, for $M$ large enough, let $B_{j}=B_{\epsilon _{1}\dots \epsilon _{j}}$%
, and $B_{1}\subset B_{2}\dots \subset B_{k}$.\newline
\noindent If $C(B_{k})>0$, 
\begin{align*}
\text{E}(\mathcal{P}_{M}(B_{k})|\boldsymbol{Y})& =\prod_{j=1}^{k}\frac{%
\alpha _{\epsilon _{1}\dots \epsilon _{j}}+n_{\epsilon _{1}\dots \epsilon
_{j}}}{\sum_{i=0}^{3}(\alpha _{\epsilon _{1}\dots \epsilon
_{j-1}i}+n_{\epsilon _{1}\dots \epsilon _{j-1}i})}=\prod_{j=1}^{k}\frac{%
\frac{j^{2}}{N}+C(B_{j})+O(\frac{1}{\sqrt{N}})}{\frac{4j^{2}}{N}+C(B_{{j-1}%
})+O(\frac{1}{\sqrt{N}})} \\
& =\prod_{j=1}^{k}\left\{ \frac{C(B_{j})}{C(B_{j-1})}+\frac{j^{2}-4j^{2}%
\frac{C(B_{j})}{C(B_{j-1})}+O(\sqrt{N})}{4j^{2}+n_{j-1}}\right\} \leq C(B_{k})+\prod_{j=1}^{k}\left\{ 1+\frac{3j^{2}+O(\sqrt{N})}{%
4j^{2}+n_{j-1}}\right\} -1 \\
& =C(B_{k})+\exp \left\{ \sum_{j=1}^{k}\frac{3j^{2}+O(\sqrt{N})}{%
4j^{2}+n_{j-1}}+O(\sum_{j=1}^{k}(\frac{3j^{2}+O(\sqrt{N})}{4j^{2}+n_{j-1}}%
)^{2})\right\} -1 \\
& =C(B_{k})+O\left( \sum_{j=1}^{k}\frac{3j^{2}+O(\sqrt{N})}{4j^{2}+n_{j-1}}%
\right) =C(B_{k})+O\left( \sum_{j=1}^{k}\frac{3j^{2}+O(\sqrt{N})}{%
4j^{2}+NC(B_{j-1})+O(\sqrt{N})}\right)  \\
& \leq C(B_{k})+O(\sum_{j=1}^{k}\frac{3j^{2}+O(\sqrt{N})}{NC(B_{j-1})}%
)=C(B_{k})+\max \{O(\frac{M}{\sqrt{N}}),O(\frac{M^{3}}{N})\}.
\end{align*}%
\noindent If $C(B_{k})=0$, suppose $l=\max_{i<k}\{C(B_{\epsilon _{1}\dots
\epsilon _{i}})>0\}$, 
\begin{align*}
\text{E}(\mathcal{P}_{M}(B_{k})|\boldsymbol{Y})& =\prod_{j=1}^{k}\frac{%
\alpha _{\epsilon _{1}\dots \epsilon _{j}}+n_{\epsilon _{1}\dots \epsilon
_{j}}}{\sum_{i=0}^{3}(\alpha _{\epsilon _{1}\dots \epsilon
_{j-1}i}+n_{\epsilon _{1}\dots \epsilon _{j-1}i})}=\prod_{j=1}^{l+1}\frac{%
\frac{j^{2}}{N}+C(B_{j})+O(\frac{1}{\sqrt{N}})}{\frac{4j^{2}}{N}+C(B_{{j-1}%
})+O(\frac{1}{\sqrt{N}})}\left( \frac{1}{4}\right) ^{M-l-1} \\
& \leq C(B_{l+1})(\frac{1}{4})^{M-l-1}+\max \{O(\frac{M}{\sqrt{N}}),O(\frac{%
M^{3}}{N})\}\left( \frac{1}{4}\right) ^{M-l-1} =0+\max \{O(\frac{M}{\sqrt{N}}),O(\frac{M^{3}}{N})\}.
\end{align*}%
\iffalse%
\begin{align*}
\text{E}(\mathcal{P}_{M}(B_{k})|\boldsymbol{Y})& =\prod_{j=1}^{k}\frac{%
\alpha _{\epsilon _{1}\dots \epsilon _{j}}+n_{\epsilon _{1}\dots \epsilon
_{j}}}{\sum_{i=0}^{3}(\alpha _{\epsilon _{1}\dots \epsilon
_{j-1}i}+n_{\epsilon _{1}\dots \epsilon _{j-1}i})}=\prod_{j=1}^{k}\{\frac{%
n_{j}}{n_{j-1}}+\frac{j^{2}-4j^{2}n_{j}/n_{j-1}}{4j^{2}+n_{j-1}}\} \\
& \leq \frac{n_{k}}{N}+\prod_{j=1}^{k}\{1+\frac{3j^{2}}{4j^{2}+n_{j-1}}\}-1
\\
& =C(B_{k})+O(\frac{1}{\sqrt{N}})+\exp \left\{ \sum_{j=1}^{k}\frac{3j^{2}}{%
4j^{2}+n_{j-1}}+O(\sum_{j=1}^{k}(\frac{3j^{2}}{4j^{2}+n_{j-1}})^{2})\right\}
-1 \\
& =C(B_{k})+O(\frac{1}{\sqrt{N}})+O(\sum_{j=1}^{k}\frac{3j^{2}}{%
4j^{2}+n_{j-1}}) \\
& =C(B_{k})+O(\frac{1}{\sqrt{N}})+O(\sum_{j=1}^{k}\frac{3j^{2}}{%
4j^{2}+NC(B_{j})+O(\sqrt{N})}) \\
& \leq C(B_{k})+O(\frac{1}{\sqrt{N}})+O(\sum_{j=1}^{k}\frac{3j^{2}}{NC(B_{k})%
}) \\
& =C(B_{k})+\max \{O(\frac{1}{\sqrt{N}}),O(\frac{M^{3}}{\sqrt{N}})\};
\end{align*}%
\fi%
\begin{align*}
\text{var}(\mathcal{P}_{M}(B_{k})|\boldsymbol{Y})& =\text{var}%
(\prod_{j=1}^{k}Z_{\epsilon _{i}\dots \epsilon _{j}}|Y)=\text{var}%
(\prod_{j=1}^{k}Z_{j}|Y)  =\text{E}(\text{var}(Z_{1}|Y)\prod_{j=2}^{k}Z_{j}^{2}|Y)+\text{var}(\text{E%
}(Z_{1}|Y)\prod_{j=2}^{k}Z_{j}|Y) \\
& =\text{var}(Z_{1}|Y)\prod_{j=2}^{k}\text{var}(Z_{j}^{2}|Y)+E^{2}(Z_{1}|Y)%
\text{var}(\prod_{j=2}^{k}Z_{j}|Y) \\
& \leq \text{var}(Z_{1}|Y)+\text{var}(\prod_{j=2}^{k}Z_{j}|Y)\leq
\sum_{j=1}^{k}\text{var}(Z_{j}|Y) \\
& =\sum_{j=1}^{M}\frac{(\alpha _{\epsilon _{1}\dots \epsilon
_{j}}+n_{\epsilon _{1}\dots \epsilon _{j}})\{\sum_{i\neq j}(\alpha
_{\epsilon _{1}\dots \epsilon _{j-1}i}+n_{\epsilon _{1}\dots \epsilon
_{j-1}i})\}}{\{\sum_{i=0}^{3}(\alpha _{\epsilon _{1}\dots \epsilon
_{j-1}i}+n_{\epsilon _{1}\dots \epsilon
_{j-1}i})\}^{2}\{\sum_{i=0}^{3}(\alpha _{\epsilon _{1}\dots \epsilon
_{j-1}i}+n_{\epsilon _{1}\dots \epsilon _{j-1}i})+1\}} \\
& \leq \sum_{j=1}^{M}\frac{1}{\{4j^{2}+n_{j-1}+1\}}\leq \frac{M}{NC(B_{k})}%
=O\left( \frac{M}{N}\right) .
\end{align*}%
Thereby for any measurable set $B\subset I$, 
\begin{align*}
\text{E}(\mathcal{P}_{M}(B)|Y_{N})& \rightarrow C(B),\quad \text{var}(%
\mathcal{P}_{M}(B)|Y_{N})\rightarrow 0, \\
P(|\mathcal{P}_{M|Y}(B)-C(B)|\geq \epsilon )& \leq \frac{E^{2}(\mathcal{P}%
_{M|Y}(B)-C(B))+\text{var}(\mathcal{P}_{M}(B)|Y)}{\epsilon ^{2}}\rightarrow
0.
\end{align*}

\subsection{Proof of Proposition \protect\ref{prop:unif convg}}

For any $B_k=B_{\epsilon_1\dots\epsilon_k}$, $k\geq M$, 
$
\text{E}(\mathcal{P}_M(B_{k})|\boldsymbol{Y})=\prod_{j=1}^M\frac{%
\alpha_{\epsilon_1\dots\epsilon_j}+n_{\epsilon_1\dots\epsilon_j}}{%
\sum_{i=0}^3(\alpha_{\epsilon_1\dots\epsilon_{j-1}i}+n_{\epsilon_1\dots%
\epsilon_{j-1}i})}\prod_{j=M+1}^{k}\frac{1}{4}.
$
If $C(B_{\epsilon_1\dots\epsilon_k})>0$: 
\begin{align*}
\text{E}(\mathcal{P}_M(B_{k})|\boldsymbol{Y})&=(\frac{1}{4}%
)^{k-M}\prod_{j=1}^M\frac{j^2+n_{j}}{4j^2+n_{{j-1}}} \leq(\frac{1}{4})^{k-M}( C(B_M)+O(\sum_{j=1}^k\frac{3j^2+O(\sqrt{N})}{%
NC(B_{j-1})})) \\
&\leq(\frac{1}{4})^{k-M}(C(B_M)+\max\{O(\frac{M}{\sqrt{N}\gamma(M)}),O(\frac{%
M^3}{N\gamma(M)})\}).
\end{align*}
For $C\in C^1([0,1]\times[0,1])$: 
\begin{align*}
\sup|\text{E}(\mathcal{P}_M(B_{k})|\boldsymbol{Y})-C(B_{k})|&\leq(\frac{1}{4}%
)^k\sup|c(b_{\epsilon_1\dots\epsilon_k})-c(b_{\epsilon_1\dots\epsilon_M})|+%
\max\{O(\frac{M}{\sqrt{N}\gamma(M)}),O(\frac{M^3}{N\gamma(M)})\} \\
&\leq (\frac{1}{4})^k(1/2)^k\sup{|c^{\prime }|}+ \max\{O(\frac{M}{\sqrt{N}%
\gamma(M)}),O(\frac{M^3}{N\gamma(M)})\} \\
&=\max\{O(\frac{M}{\sqrt{N}\gamma(M)}),O(\frac{M^3}{N\gamma(M)})\}.
\end{align*}
If $C(B_{k})=0$, suppose $l=\max_{i<k}\{C(B_{\epsilon_1\dots\epsilon_i})>0\}$%
: 
\begin{align*}
\sup \text{E}(\mathcal{P}_M(B_{k})|\boldsymbol{Y})&=\sup(\frac{1}{4}%
)^{k-l-1}\prod_{j=1}^{l+1}\frac{\frac{j^2}{N}+C(B_{j})+O(\frac{1}{\sqrt{N}})%
}{\frac{4j^2}{N}+C(B_{{j-1}})+O(\frac{1}{\sqrt{N}})} \leq \sup(\frac{1}{4})^{k-l-1}( C(B_{l+1})+O(\sum_{j=1}^k\frac{3j^2+O(\sqrt{%
N})}{NC(B_{j-1})})) \\
&=0+\max\{O(\frac{M}{\sqrt{N}\gamma(M)}),O(\frac{M^3}{N\gamma(M)})\}.
\end{align*}
Thereby $\sup_{B}|\text{E}(\mathcal{P}_M|\boldsymbol{Y})-C|\rightarrow 0$.

By the proof of Proposition \ref{prop: pw convg}, 
\begin{align*}
\sup \text{var}(\mathcal{P}_M(B_\epsilon)|\boldsymbol{Y})&\leq \sup O(\frac{M%
}{NC(B_k)})\leq O(\frac{M}{N\gamma(M)}).
\end{align*}

Let $S_M^{\delta}=\{B_{\epsilon_1\dots\epsilon_M}: \exists x\in
B_{\epsilon_1\dots\epsilon_M}, c(x)<\delta \}$, $J_M^\delta=\cup_{B\in
S_M^\delta}B$, thereby $\inf_{I/J_M^{\delta}}c(x)\geq\delta$. By $C \in
C^1(I)$, $\forall \epsilon>0$, for $M$ large enough, $\forall
B\in\{B_{\epsilon_1\dots\epsilon_M}\}, x,y\in B$, $|c(x)-c(y)|\leq
\epsilon/8 $, taking $\delta=\epsilon/4$, $\epsilon/4>\sup_{J_M^{%
\epsilon/8}}c(x)$. Therefore, 
\begin{align*}
d_{TV}(\mathcal{P}_{M|Y},C)&=\int_I|p_{M|Y}(x)-c(x)|dx =\int_{I/J_M^{\epsilon/8}}|p_{M|Y}(x)-c(x)|dx+\int_{J_M^{%
\epsilon/8}}|p_{M|Y}(x)-c(x)|dx=I_1+I_2,
\end{align*}
where $p_{M|Y}$ is the density function of $\mathcal{P}_{M|Y}$.\begin{align*}
I_2&=\int_{J_M^{\epsilon/8}}|p_{M|Y}(x)-c(x)|dx \leq
\int_{J_M^{\epsilon/8}}p_{M|Y}(x)dx+\int_{J_M^{\epsilon/8}}c(x)dx \leq
\int_{I/J_M^{\epsilon/8}}|p_{M|Y}(x)-c(x)|dx+2\int_{J_M^{\epsilon/8}}c(x)dx.
\end{align*}
\begin{align*}
d_{TV}(\mathcal{P}_{M|Y},C)&\leq 2
\int_{I/J_M^{\epsilon/8}}|p_{M|Y}(x)-c(x)|dx +2\int_{J_M^{\epsilon/8}}c(x)dx
\leq 2 \int_{I/J_M^{\epsilon/8}}|p_{M|Y}(x)-c(x)|dx +\dot\epsilon/2.
\end{align*}
\begin{align*}
I_1&=\int_{I/J_M^{\epsilon/8}}|p_{M|Y}(x)-c(x)|dx=\int_{I/J_M^{%
\epsilon/8}}|2^{2M}P_{M|Y}(B_x)-c(x)+c(b_x)-c(b_x)|dx \\
&\leq\sum_{B\in \{B_{\epsilon_1\dots\epsilon_M}\}/S_M^{\epsilon/8}}\left\{|%
\mathcal{P}_{M|Y}(B)-C(B)|+\int_{B}|c(b)-c(x)|dx\right\} \\
\end{align*}
where $B_x\in \{B_{\epsilon_1\dots\epsilon_M}\}$ such that $x\in B_x$, and $%
C(B_x)=c(b_x)\mu(B_x)$. $\forall B\in \{B_{\epsilon_1\dots\epsilon_M}\}$, $%
b,x\in B$, $|c(b)-c(x)|\leq\epsilon/8$,  we have
$
\sum_{{\{B_{\epsilon_1\dots\epsilon_M}\}/S_M^{\epsilon/8}}%
}\left\{\int_{B}|c(b)-c(x)|dx\right\}\leq \epsilon/8.
$

\begin{align*}
P& \left( \sum_{\{B_{\epsilon _{1}\dots \epsilon _{M}}\}/S_{M}^{\epsilon
/8}}|\mathcal{P}_{M|Y}(B)-C(B)|>\epsilon /4\right) \leq P\left( \max_{B\in
\{B_{\epsilon _{1}\dots \epsilon _{M}}\}/S_{M}^{\epsilon /8}}|\mathcal{P}%
_{M|Y}(B)-C(B)|\geq \frac{\epsilon }{2^{2M+2}}\right) \\
& \leq \sum_{B\in \{B_{\epsilon _{1}\dots \epsilon _{M}}\}/S_{M}^{\epsilon
/8}}P(|\mathcal{P}_{M|Y}(B)-C(B)|\geq \frac{\epsilon }{2^{2M+2}}) \\
& \leq 2^{2M+2}(\frac{2^{2M}}{\epsilon })^{2}\left\{ \sup |\text{E}(\mathcal{%
P}_{M|Y}(B))-C(B)|^{2}+\sup \text{var}(\mathcal{P}_{M|Y}(B))\right\} \\
& =2^{6M}\max \{O(\frac{M}{\sqrt{N}\gamma (M)})^{2},O(\frac{M^{3}}{N\gamma
(M)})^{2},O(\frac{M}{N\gamma (M)})\}.
\end{align*}%
Note that here 
$
r(M)\sim \min_{\{B_{\epsilon _{1}\dots \epsilon _{M}}\}/S_{M}^{\epsilon
/8}}C(B_{M})\geq \epsilon /2^{2M+2}.
$
Thus, by taking $N\propto O(2^{10M}M^{2+\eta })$, 
$
P(d_{TV}(\mathcal{P}_{M},C)\geq \epsilon |Y)=O(\frac{1}{M^{\eta }}%
)\rightarrow 0.
$

\subsection{Proof of Proposition \protect\ref{prop: order}}

\begin{enumerate}
\item For $c\geq\xi>0$, $\forall B_M$, $\exists b_M\in I$, such that $%
C(B_M)=2^{-2M}c(b_M)\geq 2^{-2M}\xi$, thereby $\gamma \sim 2^{-2M}$, %$N\propto O(\gamma^{-2}(M)M^{2+\eta})$.

\item We assume $\rho <0$, let $\alpha =\Phi ^{-1}(2^{-M})$, by symmetry of
Gaussian copula, for fixed $\rho $,
\begin{align*}
\gamma (M)& =\int_{-\infty }^{\alpha }\int_{-\infty }^{\alpha }\frac{1}{2\pi 
\sqrt{1-\rho ^{2}}}\exp \left\{ -\frac{x^{2}+y^{2}-2\rho xy}{2(1-\rho ^{2})}%
\right\} dx\,dy \\
& \geq \int_{-\infty }^{\alpha }\int_{-\infty }^{\alpha }\frac{1}{2\pi \sqrt{%
1-\rho ^{2}}}\exp \left\{ -\frac{(1-\rho )(x^{2}+y^{2})}{2(1-\rho ^{2})}%
\right\} dx\,dy =\Phi ^{2}(\alpha \sqrt{1+\rho })\sqrt{\frac{1+\rho }{1-\rho }}\approx
2^{-2M}.
\end{align*}%
\end{enumerate}

\bibliography{copula}

\newcommand{\noop}[1]{}
\begin{thebibliography}{29}
\newcommand{\enquote}[1]{``#1''}
\expandafter\ifx\csname natexlab\endcsname\relax\def\natexlab#1{#1}\fi

\bibitem[{Azzalini and Capitanio(1999)}]{azzalini1999statistical}
Azzalini, A. and Capitanio, A. (1999), \enquote{Statistical Applications of the
  Multivariate Skew Normal Distribution,} \textit{Journal of the Royal
  Statistical Society: Series B (Statistical Methodology)}, 61, 579--602.

\bibitem[{Behnen et~al.(1985)Behnen, Hu{\v{s}}kov{\'a}, and
  Neuhaus}]{behnen1985rank}
Behnen, K., Hu{\v{s}}kov{\'a}, M., and Neuhaus, G. (1985), \enquote{Rank
  Estimators of Scores for Testing Independence,} \textit{Statistics \& Risk
  Modeling}, 3, 239--262.

\bibitem[{Chen and Huang(2007)}]{chen2007nonparametric}
Chen, S.~X. and Huang, T.-M. (2007), \enquote{Nonparametric Estimation of
  Copula Functions for Dependence Modelling,} \textit{Canadian Journal of
  Statistics}, 35, 265--282.

\bibitem[{Deheuvels(1979)}]{deheuvels1979fonction}
Deheuvels, P. (1979), \enquote{La Fonction de D{\'e}pendance Empirique et Ses
  Propri{\'e}t{\'e}s. Un Test Non Param{\'e}trique d'Ind{\'e}pendance,}
  \textit{Acad{\'e}mie Royale de. Belgique. Bulletin de la Classe des Sciences.
  6e S{\'e}rie.}, 65, 274--292.

\bibitem[{Devroye and Gy{\"o}rfi(1985)}]{devroye1985nonparametric}
Devroye, L. and Gy{\"o}rfi, L. (1985), \textit{Nonparametric Density
  Estimation: the L1 View}, vol. 119 of \textit{Wiley Series in Probability and
  Statistics}, New York, NY: Wiley.

\bibitem[{Dortet-Bernadet(2005)}]{dortetbayesian}
Dortet-Bernadet, J.-L. (2005), \enquote{Bayesian Inference on Copulas and Tests
  of Independence,} Unpublished manuscript.

\bibitem[{Ferguson(1974)}]{ferguson1974prior}
Ferguson, T.~S. (1974), \enquote{Prior Distributions on Spaces of Probability
  Measures,} \textit{The Annals of Statistics}, 2, 615--629.

\bibitem[{Filippi and Holmes(\noop{3001}in press)}]{filippi2016bayesian}
Filippi, S. and Holmes, C.~C. (\noop{3001}in press), \enquote{A Bayesian
  Nonparametric Approach to Testing for Dependence Between Random Variables,}
  \textit{Bayesian Analysis}.

\bibitem[{Gasser and M{\"u}ller(1979)}]{gasser1979kernel}
Gasser, T. and M{\"u}ller, H.-G. (1979), \textit{Smoothing Techniques for Curve
  Estimation}, Heidelberg, Germany: Springer, chap. Kernel Estimation of
  Regression Functions, pp. 23--68.

\bibitem[{Genest et~al.(1995)Genest, Ghoudi, and
  Rivest}]{genest1995semiparametric}
Genest, C., Ghoudi, K., and Rivest, L.-P. (1995), \enquote{A Semiparametric
  Estimation Procedure of Dependence Parameters in Multivariate Families of
  Distributions,} \textit{Biometrika}, 82, 543--552.

\bibitem[{Gijbels and Mielniczuk(1990)}]{gijbels1990estimating}
Gijbels, I. and Mielniczuk, J. (1990), \enquote{Estimating the Density of a
  Copula Function,} \textit{Communications in Statistics-Theory and Methods},
  19, 445--464.

\bibitem[{Hanson(2006)}]{hanson2012inference}
Hanson, T.~E. (2006), \enquote{Inference for Mixtures of Finite P{\'o}lya Tree
  Models,} \textit{Journal of the American Statistical Association}, 101,
  1548--1565.

\bibitem[{Hominal and Deheuvels(1979)}]{hominal1979estimation}
Hominal, P. and Deheuvels, P. (1979), \enquote{Estimation Non Param{\'e}trique
  de la Densit{\'e} Compte-tenu d'Informations sur le Support,} \textit{Revue
  de Statistique Appliqu{\'e}e}, 27, 47--68.

\bibitem[{Jaworski et~al.(2010)Jaworski, Durante, Hardle, and
  Rychlik}]{jaworski2010copula}
Jaworski, P., Durante, F., Hardle, W.~K., and Rychlik, T. (2010),
  \textit{Copula Theory and Its Applications}, Heidelberg, Germany: Springer.

\bibitem[{Joe(1997)}]{joe1997multivariate}
Joe, H. (1997), \textit{Multivariate Models and Multivariate Dependence
  Concepts}, Boca Raton, FL: CRC Press.

\bibitem[{John(1984)}]{john1984boundary}
John, R. (1984), \enquote{Boundary Modification for Kernel Regression,}
  \textit{Communications in Statistics-Theory and Methods}, 13, 893--900.

\bibitem[{Lavine(1992)}]{lavine1992some}
Lavine, M. (1992), \enquote{Some Aspects of {P}{\'o}lya Tree Distributions for
  Statistical Modelling,} \textit{The Annals of Statistics}, 20, 1222--1235.

\bibitem[{M{\"u}ller(1991)}]{muller1991smooth}
M{\"u}ller, H.-G. (1991), \enquote{Smooth Optimum Kernel Estimators Near
  Endpoints,} \textit{Biometrika}, 78, 521--530.

\bibitem[{Nelsen(2007)}]{nelsen2007introduction}
Nelsen, R.~B. (2007), \textit{An Introduction to Copulas}, New York, NY:
  Springer.

\bibitem[{Oakes(1982)}]{oakes1982model}
Oakes, D. (1982), \enquote{A Model for Association in Bivariate Survival Data,}
  \textit{Journal of the Royal Statistical Society. Series B (Methodological)},
  44, 414--422.

\bibitem[{Oakes(1986)}]{oakes1986semiparametric}
--- (1986), \enquote{Semiparametric Inference in a Model for Association in
  Bivanate Survival Data,} \textit{Biometrika}, 73, 353--361.

\bibitem[{Paddock et~al.(2003)Paddock, Ruggeri, Lavine, and
  West}]{paddock2003randomized}
Paddock, S.~M., Ruggeri, F., Lavine, M., and West, M. (2003),
  \enquote{Randomized Polya tree models for nonparametric Bayesian inference,}
  \textit{Statistica Sinica}, 13, 443--460.

\bibitem[{Scaillet et~al.(2007)Scaillet, Charpentier, and
  Fermanian}]{charpentier2007estimation}
Scaillet, O., Charpentier, A., and Fermanian, J.-D. (2007), \enquote{The
  Estimation of Copulas: Theory and Practice,} \textit{Copulas: from Theory to
  Applications in Finance}, 35--62.

\bibitem[{Schervish(1995)}]{schervish1995theory}
Schervish, M.~J. (1995), \textit{Theory of Statistics}, New York, NY: Springer.

\bibitem[{Schuster(1985)}]{schuster1985incorporating}
Schuster, E.~F. (1985), \enquote{Incorporating Support Constraints into
  Nonparametric Estimators of Densities,} \textit{Communications in
  Statistics-Theory and Methods}, 14, 1123--1136.

\bibitem[{Schweizer and Wolff(1981)}]{schweizer1981nonparametric}
Schweizer, B. and Wolff, E.~F. (1981), \enquote{On Nonparametric Measures of
  Dependence for Random Variables,} \textit{The Annals of Statistics}, 9,
  879--885.

\bibitem[{Sklar(1959)}]{sklar1959fonctions}
Sklar, A. (1959), \textit{Fonctions de R{\'e}partition {\`a} n Dimensions et
  Leurs Marges}, Paris, France: Universit{\'e} Paris 8.

\bibitem[{Wong et~al.(2010)Wong, Ma, et~al.}]{wong2010optional}
Wong, W.~H., Ma, L., et~al. (2010), \enquote{Optional {P}{\'o}lya Tree and
  {B}ayesian Inference,} \textit{The Annals of Statistics}, 38, 1433--1459.

\bibitem[{Wu et~al.(2014)Wu, Wang, and Walker}]{wu2013bayesian}
Wu, J., Wang, X., and Walker, S.~G. (2014), \enquote{Bayesian Nonparametric
  Inference for a Multivariate Copula Function,} \textit{Methodology and
  Computing in Applied Probability}, 16, 747--763.

\end{thebibliography}

\newpage
\setcounter{page}{1}
\begin{center}
\bf \LARGE
\uppercase {Supplementary Material}
\end{center}
\bigskip 

\setcounter{equation}{0}
\setcounter{section}{0}
\setcounter{figure}{0}
\setcounter{example}{0}
\setcounter{theorem}{0}
\setcounter{table}{0}

\renewcommand {\theexample} {S.\arabic{example}}
\renewcommand {\thefigure} {S.\arabic{figure}}
\renewcommand {\thetable} {S.\arabic{table}}
\renewcommand {\theequation} {S.\arabic{equation}}
\renewcommand {\thelemma} {S.\arabic{lemma}}
\renewcommand {\thesection} {S.\arabic{section}}
\renewcommand {\thetheorem} {S.\arabic{theorem}}
\renewcommand {\thecorollary} {S.\arabic{corollary}}

\section{Common copulas}

\label{append:copula}

\subsection{Gaussian Copula}

The copula density of a bivariate Gaussian copula is given by 
\begin{align}
c(u,v)=\frac{1}{\sqrt{1-\rho^2}}exp\left\{-\frac{\rho^2(x^2+y^2)-2\rho xy}{%
2(1-\rho^2)}\right\},
\end{align}
where $\rho\in[-1,1]$ is the correlation parameter of the copula, $%
x=\Phi^{-1}(u)$, $y=\Phi^{-1}(v)$, and $\Phi^{-1}$ is the inverse of the
standard univariate Gaussian CDF.

%%student t copula

\subsection{Student's t Copula}

The copula density of a bivariate Student's t-copula follows 
\begin{align}
c(u,v)=\frac{\Gamma(\frac{\nu+2}{2})/\Gamma(\frac{\nu}{2})}{\nu\pi
f_{t_\nu}(x)f_{t_\nu}(y)\sqrt{1-\rho^2}}\left\{1+\frac{x^2+y^2-2\rho xy}{%
\nu(1-\rho^2)}\right\}^{-\frac{\nu+1}{2}},
\end{align}
where the two parameters, the correlation $\rho\in[-1,1]$ and the
degree of freedom $\nu>0$, $x=F_{t_\nu}(u)$, $y=F_{t_\nu}(v)$, and $%
f_{t_\nu} $ and $F_{t_\nu}$ are the PDF and CDF of the standard univariate
Student's t-distribution with the degree of freedom of $\nu$.

%%gumbel copula

\subsection{Gumbel Copula}
\iffalse
Another classic copula is the Gumbel copula, which is a subclass of
Archimedean copulas.

\begin{mydef}
\citep{nelsen2007introduction} A (two-dimensional) copula $C$ is called
Archimedean copula if it follows: $C(u,v)=\phi ^{\lbrack -1]}(\phi (u)+\phi
(v))$, where $\phi $ is continuous, strictly decreasing function from $%
I=[0,1]$ to $[0,\infty ]$ such that $\phi (1)=0$, and $\phi ^{\lbrack -1]}$
is the pseudo-inverse of $\phi $ given by 
\begin{equation*}
\phi (\phi ^{\lbrack -1]}(t))=%
\begin{cases}
t, & 0\leq t\leq \phi (0), \\ 
\phi (0), & \phi (0)\leq t\leq \infty .%
\end{cases}%
\end{equation*}
\end{mydef}

When the generator function $\phi _{a}(t)=(-log(t))^{a}$, the Archimedean
copula generated is called Gumbel copula $C(u,v)=\exp \{-[\{-\log
(u)\}^{a}+\{-\log (v)\}^{a}]^{1/a}\}$, with density: 
\fi
The copula density of a bivariate Gumbel copula is given by
\begin{align*}
c(u,v)& =C(u,v)(uv)^{-1}[\{-\log (u)\}^{a}+\{-\log (v)\}^{a}]^{-2+2/a}\{\log
(u)\log (v)\}^{a-1} \\
& (1+(a-1)[\{-\log (u)\}^{a}+\{-\log (v)\}^{a}]^{-1/a}),
\end{align*}%
where $a\geq 1$ is the dependence parameter. %%skew-normal

\subsection{Skew-normal Copula}

A d-dimensional random vector $\boldsymbol{Z}=(Z_{1},\dots ,Z_{d})^{T}$
follows a skew-normal distribution \citep{azzalini1999statistical}, denoted $%
\boldsymbol{Z}\sim SN_{d}(\Omega ,\boldsymbol{\alpha })$ if 
\begin{equation*}
\boldsymbol{Z}=%
\begin{cases}
\boldsymbol{X} & \text{if }\,X_{0}>0, \\ 
-\boldsymbol{X} & \text{otherwise,}%
\end{cases}%
\end{equation*}%
where $(X_{0},\boldsymbol{X})^{T}\sim N_{d+1}(\boldsymbol{0},\Omega ^{\ast })
$, $\Omega ^{\ast }=%
\begin{pmatrix}
1 & \boldsymbol{\delta }^{T} \\ 
\boldsymbol{\delta } & \Omega 
\end{pmatrix}%
$ and $\boldsymbol{\alpha }=\frac{1}{(1-\boldsymbol{\delta }^{T}\Omega ^{-1}%
\boldsymbol{\delta })^{1/2}}\Omega ^{-1}\boldsymbol{\delta }$. And the
density functions for $\boldsymbol{Z}$ is $f_{SN,d}(\boldsymbol{z};\Omega ,%
\boldsymbol{\alpha })=2\phi _{d}(\boldsymbol{z};\Omega )\Phi (\boldsymbol{%
\alpha }^{T}\boldsymbol{z})$, where $\phi _{d}(\cdot ;\Omega )$ is the $d$%
-dimensional normal density with zero mean and correlation matrix $\Omega $.

For the marginals, suppose $\boldsymbol{Z}$ is partitioned as $\boldsymbol{Z}%
=(\boldsymbol{Z}_{1}^{T},\boldsymbol{Z}_{2}^{T})^{T}$ of dimension $h$ and $%
d-h$; $\Omega $ and $\boldsymbol{\alpha }$ by  
\begin{equation*}
\Omega =%
\begin{pmatrix}
\Omega _{11} & \Omega _{12} \\ 
\Omega _{21} & \Omega _{22}%
\end{pmatrix}%
,\,\boldsymbol{\alpha }=%
\begin{pmatrix}
\boldsymbol{\alpha }_{1} \\ 
\boldsymbol{\alpha }_{2}%
\end{pmatrix}%
,
\end{equation*}%
then the marginal distribution of $\boldsymbol{Z}_{1}$ is $SN_{d}(\Omega
_{11},\boldsymbol{\bar{\alpha}}_{1})$, where 
\begin{equation*}
\bar{\boldsymbol{\alpha }}_{1}=\frac{\boldsymbol{\alpha }_{1}+\Omega
_{11}^{-1}\Omega _{12}\boldsymbol{\alpha }_{2}}{(1+\boldsymbol{\alpha }%
^{T}\Omega _{22\cdot 1}\boldsymbol{\alpha }_{2})^{1/2}},\,\Omega _{22\cdot
1}=\Omega _{22}-\Omega _{21}\Omega _{21}^{-1}\Omega _{12}.
\end{equation*}

So the bivariate skew-normal copula density is 
\begin{align}
c(u,v)=\frac{\phi_2((x,y)^T,\Omega)\Phi(\alpha_1x+\alpha_2y)}{2\phi(x)\Phi(%
\bar{\alpha}_1x)\phi(y)\Phi(\bar\alpha_2y)},
\end{align}
where $\Omega=%
\begin{pmatrix}
1 & \rho \\ 
\rho & 1%
\end{pmatrix}%
$; $\rho\in[-1,1]$, $\alpha_1$,$\alpha_2$ are parameters, $x=F_{SN_1, \bar{%
\alpha}_1}^{-1}(u)$, $y=F_{SN_1, \bar{\alpha}_2}^{-1}(v)$, and $F_{SN_1,
\alpha}$ is the CDF of $Z\sim SN_1(1,\alpha)$.

\section{Additional simulation results}
%\subsection{Comprehensive comparison between D-P tree and Gaussian mixture model}
\label{append: sim} \label{append: sim} 
\begin{table}[!h]
\centering
\begin{tabular}{rrr|ll|ll|ll}
\hline\hline
& \multirow{3}{*}{$\rho$} & \multirow{ 3}{*}{$\alpha$} & \multicolumn{6}{c}{N
} \\ \cline{4-9}
&  &  & \multicolumn{2}{c}{1000} & \multicolumn{2}{c}{10,000} & 
\multicolumn{2}{c}{100,000} \\ \hline
1 & 0.50 & (2,0) & 0.12 & 0.00 & 0.07 & 0.00 & 0.04 & xx \\ 
2 & 0.90 & (2,0) & 0.18 & 0.01 & 0.09 & 0.01 & 0.04 & xx \\ 
3 & 0.50 & (-10,50) & 0.18 & 0.07 & 0.09 & 0.06 & 0.05 & xx \\ 
4 & 0.90 & (-10,50) & 0.21 & 0.03 & 0.10 & 0.03 & 0.05 & xx \\ 
5 & 0.50 & (50,0) & 0.12 & 0.01 & 0.07 & 0.00 & 0.04 & xx \\ 
6 & 0.90 & (50,0) & 0.16 & 0.05 & 0.08 & 0.05 & 0.04 & xx \\ 
7 & 0.50 & (100,-100) & 0.26 & 0.17 & 0.14 & 0.17 & 0.07 & 0.17 \\ 
8 & 0.90 & (100,-100) & 0.46 & 0.16 & 0.20 & 0.16 & 0.08 & 0.16 \\ 
\hline\hline
\end{tabular}%
\caption{Comparison of the K-L divergence between the D-P tree (left) and
the Gaussian mixture (right) estimation for skew-normal target copulas.}\label{table: mixnorm all}
\end{table}
%\subsection{Comprehensive results for comparison between the D-P tree and the frequentist methods}
\begin{table}[!h]
\centering
\begin{tabular}{r|llll|llll}
\hline\hline
$N$ & D-P Tree & Empirical & Kernel & Hist, & D-P Tree & Empirical & Kernel
& Hist. \\ \hline
 & \multicolumn{4}{c|}{$\rho=0.5, \alpha=(-10,50)$} & \multicolumn{4}{c}{$\rho=0.5, \alpha=(100,-100)$} \\ \hline
10 & 0.242 & NA & 0.337 & Inf & 0.528 & NA & 0.528 & Inf \\ 
20 & 0.223 & NA & 0.247 & Inf & 0.473 & NA & 0.428 & Inf \\ 
50 & 0.178 & NA & 0.165 & Inf & 0.386 & NA & 0.314 & Inf \\ 
100 & 0.161 & NA & 0.123 & Inf & 0.349 & NA & 0.261 & Inf \\ 
500 & 0.098 & NA & 0.064 & Inf & 0.222 & NA & 0.166 & Inf \\ 
1,000 & 0.085 & NA & 0.047 & Inf & 0.184 & NA & 0.136 & Inf \\ 
5,000 & 0.056 & NA & 0.027 & Inf & 0.112 & NA & 0.090 & Inf \\ 
10,000 & 0.041 & NA & 0.020 & Inf & 0.089 & NA & 0.076 & Inf \\ \hline
\hline
 & \multicolumn{4}{c|}{$\rho=0.9, \alpha=(-10,50)$} & \multicolumn{4}{c}{$\rho=0.9, \alpha=(100,-100)$} \\ \hline
10 & 0.441 & NA & 0.449 & Inf & 1.068 & NA & 1.099 & Inf \\ 
20 & 0.38 & NA & 0.317 & Inf & 1.013 & NA & 0.969 & Inf \\ 
50 & 0.328 & NA & 0.225 & Inf & 0.836 & NA & 0.763 & Inf \\ 
100 & 0.28 & NA & 0.146 & Inf & 0.715 & NA & 0.619 & Inf \\ 
500 & 0.163 & NA & 0.065 & Inf & 0.479 & NA & 0.410 & Inf \\ 
1,000 & 0.121 & NA & 0.044 & Inf & 0.379 & NA & 0.345 & Inf \\ 
5,000 & 0.068 & NA & 0.021 & Inf & 0.209 & NA & 0.227 & Inf \\ 
10,000 & 0.051 & NA & 0.013 & Inf & 0.164 & NA & 0.191 & Inf \\ \hline\hline
\end{tabular}%
\caption{Comparison of the K-L divergence between the D-P tree posterior
mean estimator and the frequentist estimators for skew-normal target
copulas. }\label{table: kl all}
\end{table}

\begin{table}[!h]
\centering
\begin{tabular}{rllll|llll}
\hline
\hline
$N$ & D-P Tree & Empirical & Kernel & Hist. & D-P Tree & Empirical & Kernel
& Hist. \\ \hline
 & \multicolumn{4}{c|}{$\rho=0.5, \alpha=(-10,50)$} & \multicolumn{4}{c}{$\rho=0.5, \alpha=(100,-100)$} \\ \hline
10 & 0.773 & NA & 2.742 & 93.588 & 1.365 & NA & 2.190 & 71.788 \\ 
20 & 0.814 & NA & 1.926 & 55.190 & 1.163 & NA & 2.657 & 56.726 \\ 
50 & 0.728 & NA & 1.690 & 35.150 & 1.050 & NA & 1.177 & 36.757 \\ 
100 & 0.918 & NA & 1.350 & 25.870 & 1.159 & NA & 1.347 & 25.723 \\ 
500 & 0.604 & NA & 0.526 & 11.437 & 1.072 & NA & 1.398 & 11.665 \\ 
1,000 & 0.528 & NA & 0.487 & 8.133 & 0.894 & NA & 0.703 & 8.078 \\ 
5,000 & 0.525 & NA & 0.412 & 3.617 & 0.703 & NA & 0.516 & 3.601 \\ 
10,000 & 0.389 & NA & 0.263 & 2.565 & 0.701 & NA & 0.769 & 2.600 \\ \hline
\hline
 & \multicolumn{4}{c|}{$\rho=0.9, \alpha=(-10,50)$} & \multicolumn{4}{c}{$\rho=0.9, \alpha=(100,-100)$} \\ \hline
10 & 1.195 & NA & 2.365 & 74.140 & 2.281 & NA & 7.372 & 86.920 \\ 
20 & 1.297 & NA & 1.837 & 54.820 & 3.335 & NA & 6.047 & 53.928 \\ 
50 & 1.521 & NA & 1.689 & 34.064 & 2.217 & NA & 2.185 & 36.227 \\ 
100 & 1.640 & NA & 1.427 & 24.826 & 2.098 & NA & 2.380 & 25.320 \\ 
500 & 1.132 & NA & 0.951 & 11.515 & 1.980 & NA & 1.826 & 11.277 \\ 
1,000 & 0.838 & NA & 0.994 & 8.073 & 1.765 & NA & 1.535 & 8.204 \\ 
5,000 & 0.700 & NA & 0.420 & 3.643 & 1.540 & NA & 1.146 & 3.694 \\ 
10,000 & 0.597 & NA & 0.300 & 2.575 & 1.949 & NA & 1.177 & 2.843 \\ 
\hline\hline
\end{tabular}%
\caption{Comparison of the $\protect\sqrt{MISE}$ between the D-P tree
posterior mean estimator and the frequentist estimators for skew-normal
target copulas.}\label{table: mise all}
\end{table}

\begin{table}[!h]
\centering
\begin{tabular}{r|llll|llll}
\hline\hline
$N$ & D-P Tree & Empirical & Kernel & Hist. & D-P Tree & Empirical & Kernel
& Hist. \\ \hline
 & \multicolumn{4}{c|}{$\rho=0.5, \alpha=(-10,50)$} & \multicolumn{4}{c}{$\rho=0.5, \alpha=(100,-100)$} \\ \hline
10 & 0.064 & 0.120 & 0.083 & 0.120 & 0.072 & 0.118 & 0.091 & 0.117 \\ 
20 & 0.057 & 0.083 & 0.065 & 0.083 & 0.065 & 0.082 & 0.068 & 0.082 \\ 
50 & 0.044 & 0.057 & 0.048 & 0.057 & 0.044 & 0.057 & 0.050 & 0.057 \\ 
100 & 0.027 & 0.037 & 0.029 & 0.037 & 0.037 & 0.041 & 0.038 & 0.041 \\ 
500 & 0.016 & 0.018 & 0.016 & 0.018 & 0.018 & 0.017 & 0.017 & 0.017 \\ 
1,000 & 0.013 & 0.013 & 0.013 & 0.013 & 0.013 & 0.012 & 0.013 & 0.012 \\ 
5,000 & 0.006 & 0.006 & 0.006 & 0.006 & 0.007 & 0.006 & 0.007 & 0.006 \\ 
10,000 & 0.004 & 0.004 & 0.005 & 0.004 & 0.005 & 0.004 & 0.005 & 0.004 \\ 
\hline
\hline
 & \multicolumn{4}{c|}{$\rho=0.9, \alpha=(-10,50)$} & \multicolumn{4}{c}{$\rho=0.9, \alpha=(100,-100)$} \\ \hline
10 & 0.080 & 0.129 & 0.102 & 0.129 & 0.080 & 0.121 & 0.102 & 0.121 \\ 
20 & 0.065 & 0.089 & 0.072 & 0.089 & 0.075 & 0.101 & 0.089 & 0.101 \\ 
50 & 0.055 & 0.060 & 0.054 & 0.060 & 0.046 & 0.056 & 0.051 & 0.056 \\ 
100 & 0.038 & 0.038 & 0.037 & 0.038 & 0.037 & 0.036 & 0.036 & 0.036 \\ 
500 & 0.020 & 0.018 & 0.019 & 0.018 & 0.021 & 0.018 & 0.020 & 0.018 \\ 
1,000 & 0.014 & 0.014 & 0.014 & 0.014 & 0.014 & 0.013 & 0.014 & 0.013 \\ 
5,000 & 0.006 & 0.006 & 0.006 & 0.006 & 0.007 & 0.005 & 0.008 & 0.005 \\ 
10,000 & 0.005 & 0.004 & 0.005 & 0.004 & 0.005 & 0.004 & 0.006 & 0.004 \\ 
\hline\hline
\end{tabular}%
\caption{Comparison of the $\protect\sqrt{MISE_C}$ between the D-P tree
posterior mean estimator and the frequentist estimators for the skew-normal
target copulas.}\label{table: misec all}
\end{table}

\begin{table}[!h]
\centering
\begin{tabular}{r|llll|llll}
\hline\hline
 & \multicolumn{4}{c|}{$\rho=0.5, \alpha=(-10,50)$} & \multicolumn{4}{c}{$\rho=0.5, \alpha=(100,-100)$} \\ \hline
$N$ & D-P Tree & Empirical & Kernel & Hist. & D-P Tree & Empirical & Kernel
& Hist. \\ \hline
10 & 0.026 & 0.317 & 0.028 & 0.317 & 0.054 & 0.321 & 0.054 & 0.321 \\ 
20 & 0.026 & 0.225 & 0.028 & 0.225 & 0.054 & 0.230 & 0.055 & 0.230 \\ 
50 & 0.026 & 0.144 & 0.026 & 0.144 & 0.054 & 0.151 & 0.054 & 0.151 \\ 
100 & 0.026 & 0.103 & 0.026 & 0.103 & 0.054 & 0.113 & 0.054 & 0.113 \\ 
500 & 0.026 & 0.051 & 0.026 & 0.051 & 0.053 & 0.070 & 0.053 & 0.070 \\ 
1,000 & 0.026 & 0.041 & 0.026 & 0.041 & 0.053 & 0.062 & 0.053 & 0.062 \\ 
5,000 & 0.026 & 0.029 & 0.026 & 0.029 & 0.053 & 0.055 & 0.053 & 0.055 \\ 
10,000 & 0.026 & 0.027 & 0.026 & 0.027 & 0.053 & 0.054 & 0.053 & 0.054 \\ 
\hline
\end{tabular}
\begin{tabular}{r|llll|llll}
\hline
 & \multicolumn{4}{c|}{$\rho=0.9, \alpha=(-10,50)$} & \multicolumn{4}{c}{$\rho=0.9, \alpha=(100,-100)$} \\ \hline
$N$ & D-P Tree & Empirical & Kernel & Hist. & D-P Tree & Empirical & Kernel
& Hist. \\ \hline
10 & 0.010 & 0.316 & 0.015 & 0.316 & 0.063 & 0.322 & 0.065 & 0.322 \\ 
20 & 0.010 & 0.224 & 0.014 & 0.224 & 0.063 & 0.232 & 0.064 & 0.232 \\ 
50 & 0.010 & 0.142 & 0.012 & 0.142 & 0.063 & 0.155 & 0.063 & 0.155 \\ 
100 & 0.010 & 0.100 & 0.011 & 0.100 & 0.063 & 0.118 & 0.063 & 0.118 \\ 
500 & 0.010 & 0.046 & 0.009 & 0.046 & 0.063 & 0.077 & 0.063 & 0.077 \\ 
1,000 & 0.009 & 0.033 & 0.009 & 0.033 & 0.063 & 0.070 & 0.063 & 0.070 \\ 
5,000 & 0.009 & 0.017 & 0.009 & 0.017 & 0.063 & 0.064 & 0.063 & 0.064 \\ 
10,000 & 0.009 & 0.013 & 0.009 & 0.013 & 0.063 & 0.063 & 0.063 & 0.063 \\ 
\hline\hline
\end{tabular}%
\caption{Comparison of the $\protect\sqrt{MSE_g}$ between the D-P tree
posterior mean estimator and the frequentist estimators for the skew-normal
target copulas.}\label{table: mseg all}
\end{table}

\end{document}